\documentclass[11pt]{article}
\usepackage{amsmath}   
\usepackage{amsfonts}    
\usepackage{xfrac} 
\usepackage{tabu}
\newcommand {\bu} {\mbox{\boldmath $u$}}
\newcommand {\btildeu} {\mbox{\boldmath $\tilde{u}$}}

\newcommand {\bx} {\mbox{\boldmath $x$}}
\newcommand {\btildex} {\mbox{\boldmath $\tilde{x}$}}
\newcommand {\by} {\mbox{\boldmath $y$}}
\newcommand {\bz} {\mbox{\boldmath $z$}}

\newcommand{\calC}{{\cal C}}

\newcommand{\calU}{{\cal U}}

\newcommand{\calX}{{\cal X}}

\begin{document}
\thispagestyle{empty}
\title{Exact Random Coding Exponents and Universal \\
Decoders for the Asymmetric Broadcast Channel\\}
\author{\\ Ran Averbuch and Neri Merhav\\}
\maketitle
\begin{center}
The Andrew \& Erna Viterbi Faculty of Electrical Engineering \\
Technion - Israel Institute of Technology \\
Technion City, Haifa 3200004, ISRAEL \\
\{rans@campus, merhav@ee\}.technion.ac.il
\end{center}
\vspace{1.5\baselineskip}
\setlength{\baselineskip}{1.5\baselineskip}

\begin{abstract}
This work contains two main contributions concerning the
asymmetric broadcast channel. The first is an analysis of 
the exact random coding error exponents
for both users, and the second is the derivation
of universal decoders for both users. These universal decoders are
certain variants of the maximum mutual information (MMI)
universal decoder, which achieve the corresponding random coding
exponents of optimal decoding.
In addition, we introduce some lower bounds, 
which involve optimization over very few parameters, 
unlike the original, exact exponents, which involve minimizations over
auxiliary probability distributions.
Numerical results for the binary symmetric broadcast channel show improvements over 
previously derived error exponents for the same model. \\

\end{abstract}

\newpage
\section{Introduction}

One of the most elementary system configuation models in multi-user information theory 
is the broadcast channel (BC). It has been introduced in the 
early seventies of the twentieth century by Cover \cite{COVER72}, 
and since then, a vast amount of papers and books, studying different topics of 
the broadcast problem, have been published. 
Generally speaking, the BC
is a communication model, 
where a single transmitter wishes 
to communicate different messages to two or more receivers. 
The various messages may be private (i.e., aimed to one receiver only) 
or common (i.e., aimed to two or more receivers). 

Although the characterization of the capacity region of the general BC is still 
an open problem, some special cases have been solved, most notably, the
degraded BC (DBC), first presented in \cite{COVER72}.
The capacity region of the DBC, conjectured by Cover, was first proved 
to be achievable by Bergmans \cite{BERGMANS1}, 
and the converse was established by Bergmans 
\cite{BERGMANS2} and Gallager \cite{GALLAGER74}.
Another special case, which is somewhat more general than the DBC and which was first 
introduced and solved by K\"{o}rner and Marton \cite{KM77},
is the broadcast channel with degraded message sets, also known as the asymmetric broadcast channel (ABC). The direct part of their coding theorem relys on  
Bergmans' scheme, which suggested the use of an hierarchical random code: 
First generate ``cloud centers", 
which designate messages 
intended to both the receiver with
the relatively high channel quality, 
henceforth referred to as the {\it strong user}, and the receiver with
the relatively low channel quality, henceforth referred to as the {\it weak user}.
Then, in the second step, ``around" each cloud center,
generate a codeword for each message 
that is intended to the strong user
only. The transmitter sends a codeword pertaining to one of the clouds. 
The strong decoder 
fully decodes both the common message and his private message,
whereas the weak decoder decodes the common message only. 
Other channels in which one receiver is superior to another 
and channels with nested information were studied by 
Csisz\'{a}r and K\"{o}rner \cite{CK78} and by El Gamal \cite{ELGAMAL79}, 
to name a few. 

Multi--user information theory is, first and foremost, driven by the quest to
characterize capacity regions, i.e., the region of all sets of rates that
allow reliable communication (a.k.a.\ achievable rates).
A somewhat sharper performance metric concerns the exponential decay rate
(the error exponent) of the
probability of error
for each user, as a function of the coding rates 
within the interior of the capacity region.
On top of that, an interesting question concerns the trade--off between
the error exponent of the strong user and the one of the weak user, or
equivalently, the achievable region in the plane of error exponents for a
given set of coding rates. 

While the capacity regions of the DBC and the ABC have been known for many years, 
only little has been known about their reliability functions.
Earlier works on error exponents for the general DBC and ABC 
include those of Gallager \cite{GALLAGER74} and K\"{o}rner and Sgarro \cite{KS80}, respectively. 
In both works, the coding scheme of \cite{BERGMANS1} was adopted, 
but the decoder was sub--optimal. 
More recently, Kaspi and Merhav \cite{KM11} have derived
some tighter lower bounds to the reliability functions of both users
by analyzing random coding error exponents of their optimal decoders.
While their derivation
was exponentially tight at most of the steps, 
there were still some steps in \cite{KM11} where exponential
tightness might have been compromised. 
Moreover, Kaspi and Merhav have analyzed ensembles of i.i.d.\
codes, which are not as good as ensembles of fixed composition codes \cite[Section 7.3]{GAL68}. These two points give rise to the thought that there 
is room for improvement upon the results of \cite{KM11},
and indeed, such an improvement is one of the contributions of this work. In
fact, the exponential error bounds, derived in this paper, both for the strong
user and the weak one, are tight in the sense that they provide the exact
random coding exponents for the ensemble of fixed composition codes.
Moreover, the resulting expressions are much simpler and easier to calculate
than those of the best exponential bounds of 
Kaspi and Merhav (see, in
particular, the second part of \cite{KM11}). 

Interestingly, one of the ingredients 
that contributes significantly to this
simplification in the error exponent expressions, is the derivation of
{\it universal decoders} for both users, and this simplification is achieved thanks to
a simple sandwich argument, asserting that a lower bound to 
the error exponent of the universal
decoder cannot be larger than an upper bound to the error exponent of the
optimal decoder, but on the other hand, 
the latter turns out to be mathematically smaller
than or equal to the former, and so, 
by contrasting the two exponential error bounds, which must therefore be 
equivalent, the expressions are considerably simplified. 
In other words, beyond this simplification of the error
exponent bounds, there is an additional bonus, 
which is in obtaining universal decoders
for both users. These decoders achieve the same random coding error exponents 
as the corresponding
optimal decoders of the two users. Both universal decoders are certain variants of the
maximum mutual information (MMI) decoder \cite[Theorem 5.2]{CK11}, but they are
different from the earlier proposed MMI-like universal decoders for the ABC, due to
K\"{o}rner and Sgarro \cite{KS80}. For one thing, our
universal decoder for the weak user depends explicitly on the entire code, 
unlike the one
in \cite{KS80}, which depends on the cloud centers only. 

Since we rely heavily on the method of types, our exponential error bounds
have the flavor of those of Csisz\'{a}r and K\"{o}rner \cite{CK11}. While
exponentially tight, their shortcoming is that they are not easy to calculate
since they involve minimizations over auxiliary channels, and these might be
computationally painful especially for large alphabets. To alleviate
this difficulty, we also propose Gallager--style bounds \cite{GAL68}, which require
optimizations over very 
few (one or two) parameters, but the caveat is that
exponential tightness might be sacrificed.
Moreover, the Gallager--style bounds lend themselves to better intuitive
understanding on the behavior of 
the error exponents for both of the users. 
Specifically, we derive a {\it phase diagram} for the weak user, 
which fully describes the functional behavior of the bound 
in different regions of the plane of rates. 
We also demonstrate our results numerically 
for an example of the binary symmetric BC, and
compare our results to those in earlier works, showing explicitly the improvement. 

The remaining part of the paper is organized as follows. 
In Section 2, we establish notation conventions, formalize the model and the problem, and finally, review 
some preliminaries. 
In Section 3, we summarize the main theoretical results of this paper, and give some numerical results for the binary symmetric BC. 
Section 4 provides the proofs concerning the strong user in the ABC (the exact random coding error exponent and the universal decoder), and 
Section 5 contains a similar treatment for the weak user.
In Section 6, we derive lower bounds on the exact random coding error exponents, 
and in Section 7 we study them.

\section{Notation Conventions and Problem Formulation}

\subsection{Notation Conventions}
Throughout the paper, random variables will be denoted by capital letters, specific values they 
may take will be denoted by the corresponding lower case letters, and their alphabets will be 
denoted by calligraphic letters. Random vectors and their realizations will be denoted, 
respectively, by capital letters and the corresponding lower case letters, both in the bold face 
font. Their alphabets will be superscripted by their dimensions. 
For example, the random vector $\mathbf{X} = (X_{1}, \dotsc , X_{n})$, ($n$ - positive integer) 
may take a specific vector value $\bx = (x_{1}, \dotsc , x_{n})$ in $\mathcal{X}^{n}$, 
the $n$-th order Cartesian power of $\mathcal{X}$, which is the alphabet of each component of this vector. Sources and channels will be subscripted by the names of the relevant random 
variables/vectors and their conditionings, whenever applicable, 
following the standard notation conventions, e.g., $Q_{X}$, $Q_{Y|X}$, and so on. 
When there is no room for ambiguity, these subscripts will be omitted. For a generic joint 
distribution $Q_{XY} = \{Q_{XY}(x,y), x \in \mathcal{X}, y \in \mathcal{Y} \}$, which will often 
be abbreviated by $Q$, information measures will be denoted in the conventional manner, but
with a subscript $Q$, that is, $H_{Q}(X)$ 
is the marginal entropy of $X$, $H_{Q}(X|Y)$ is the conditional entropy of $X$ given $Y$, 
$I_{Q}(X;Y) = H_{Q}(X) - H_{Q}(X|Y)$ is the mutual information between $X$ and $Y$, 
and so on. The weighted divergence between 
two conditional distributions (channels), say, $Q_{Z|X}$ and $W = \{W(z|x), 
x \in \mathcal{X}, z \in \mathcal{Z} \}$, with weighting $Q_{X}$ is defined as
\begin{equation}
  D(Q_{Z|X} || W | Q_{X}) = \sum_{x \in \mathcal{X}} Q_{X}(x) 
\sum_{z \in \mathcal{Z}} Q_{Z|X}(z|x) \log \frac{Q_{Z|X}(z|x)}{W(z|x)},
\end{equation}
where logarithms, here and throughout the sequel, are taken to the natural base.
The probability of an event $\mathcal{E}$ will be denoted by 
$\mathrm{Pr} \{ \mathcal{E} \}$, and the expectation operator with respect to (w.r.t.) a 
probability distribution $P$ will be denoted by $\mathbb{E} \{\cdot \}$. 
For two positive sequences $a_{n}$ and $b_{n}$, the notation $a_{n} \doteq b_{n}$ will stand 
for equality in the exponential scale, that is, $\lim_{n \to \infty} \frac{1}{n} 
\log \frac{a_{n}}{b_{n}} = 0$. The indicator function of an event $\mathcal{E}$ 
will be denoted by $\mathcal{I} \{ \mathcal{E} \}$. 
The notation $[x]_{+}$ will stand for $\max \{0, x\}$. 

The empirical distribution of a sequence $\bx \in \mathcal{X}^{n}$, which will 
be denoted by $\hat{P}_{\bx}$, is the vector of relative frequencies, $\hat{P}_{\bx}(x)$, 
of each symbol $x \in \mathcal{X}$ in $\bx$. 
The type class of $\bx \in \mathcal{X}^{n}$, denoted $\mathcal{T}(\bx)$, 
is the set of all vectors $\bx'$ with $\hat{P}_{\bx'} = \hat{P}_{\bx}$. 
When we wish to emphasize the dependence of the type class on the empirical 
distribution $\hat{P}$, we will denote it by $\mathcal{T}(\hat{P})$. Information measures 
associated with empirical distributions will be denoted with 'hats' and will be subscripted 
by the sequences from which they are induced. For example, the entropy associated 
with $\hat{P}_{\bx}$, which is the empirical entropy of $\bx$, will be denoted 
by $\hat{H}_{\bx}(X)$. Similar conventions will apply to the joint empirical distribution, 
the joint type class, the conditional empirical distributions and the conditional type classes 
associated with pairs (and multiples) of sequences of length $n$. 
Accordingly, $\hat{P}_{\bx\by}$ would be the joint empirical distribution 
of $(\bx, \by) = \{(x_{i}, y_{i})\}_{i=1}^{n}$, $\mathcal{T}(\bx, 
\by)$ or $\mathcal{T}(\hat{P}_{\bx\by})$ will denote the joint type 
class of $(\bx, \by)$, $\mathcal{T}(\bx| \by)$ will stand for the 
conditional type class of $\bx$ given $\by$, $\hat{H}_{\bx\by}(X,Y)$ 
will designate the empirical joint entropy of $\bx$ and 
$\by$, $\hat{H}_{\bx\by}(X|Y)$ will be the empirical conditional entropy, 
$\hat{I}_{\bx\by}(X;Y)$ will denote the empirical mutual information, and so on. 
When we wish to emphasize the dependence of $\mathcal{T}(\bx| \by)$ upon 
$\by$ and the relevant empirical conditional distribution, $Q_{X|Y} = 
\hat{P}_{\bx|\by}$, we denote it by $\mathcal{T}( Q_{X|Y} | \by)$. 
Similar conventions will apply to triples of sequences, say, 
$\{(\bx, \by, \bz)\}$, etc. Likewise, when we wish to emphasize the 
dependence of empirical information measures upon a given empirical distribution given by $Q$, 
we denote them using the subscript $Q$, as described above.

\subsection{Problem Formulation}

We consider a memoryless ABC 
with a finite input alphabet $\mathcal{X}$ and finite output alphabets $\mathcal{Y}$ and $\mathcal{Z}$. 
Let $W_{1} = \{W_{1}(y|x),~x \in \mathcal{X},~y \in \mathcal{Y} \}$ 
and $W_{2} = \{W_{2}(z|x),~x \in \mathcal{X},~z \in \mathcal{Z} \}$
denote the single--letter input--output transition probability matrices,
associated with the strong user and the weak user, respectively.
When these channels are fed by an input vector $\bx \in \mathcal{X}^{n}$, 
they produce
the corresponding output vectors $\by \in \mathcal{Y}^{n}$ and $\bz \in \mathcal{Z}^{n}$, according to
\begin{align}
   W_{1}(\by|\bx)  &=  \prod_{t=1}^{n}  W_{1}(y_{t}|x_{t}), \\
   W_{2}(\bz|\bx)  &=  \prod_{t=1}^{n}  W_{2}(z_{t}|x_{t}).
\end{align}
We are interested in sending 
one out of $M_{y}M_{z}$ messages to the strong user, 
that observes $\by$, and 
one out of $M_{z}$ messages to the weak user, 
that observes $\bz$. 
Specifically, 
consider the following mechanism of random selection of an hierarchical code
for the ABC.
Let $\mathcal{U}$ be a finite alphabet, 
let $P_{U}$ be a given probability distribution on $\mathcal{U}$, 
and let $P_{X|U}$ be a given matrix of conditional probabilities of $X$ given $U$. 
We first select, independently at random, $M_{z} = e^{n R_{z}}$ $n$-vectors 
(``cloud centers''), 
$\bu_{0}, \bu_{1}, \dotsc, \bu_{M_{z}-1}$, 
all under the uniform distribution over the type class $\mathcal{T}(P_{U})$. 
Next, for each $i = 0,1, \dotsc, M_{z} - 1$, we select conditionally 
independently (given $\bu_{i}$), $M_{y} = e^{n R_{y}}$ codewords, 
$\bx_{i,0}, \bx_{i,1}, \dotsc, \bx_{i,(M_{y}-1)}$,
under the uniform distribution across the conditional type 
class $\mathcal{T}(P_{X|U}|\bu_{i})$. We denote the sub-code
$\mathcal{C}_{i} = \{ \bx_{i,0}, \bx_{i,1}, \dotsc, \bx_{i,(M_{y}-1)} \}$.
Once selected, the entire codebook $\mathcal{C}=\cup_{i=0}^{M_z-1}\calC_i$, 
together with the collection of all cloud centers, $\{ \bu_{0},
\bu_{1}, \dotsc, \bu_{M_{z}-1} \}$, are
revealed to the encoder and to both decoders. \\
\\
The optimal decoder for the strong user is given by
\begin{equation}
\label{MLstrong}
[\hat{i}(\by), \hat{j}(\by)] = \operatorname*{arg\,max}_{0 \leq i \leq M_{z}-1, 0 \leq j \leq M_{y}-1} W_{1}(\by|\bx_{i,j}) ,
\end{equation}
while the optimal decoder for the weak user (the bin index decoder) is given by
\begin{equation}
\label{MLweak}
\tilde{i}(\bz) = \operatorname*{arg\,max}_{0 \leq i \leq M_{z}-1} W_{2}(\bz|\mathcal{C}_{i})  ,
\end{equation}
where
\begin{equation}
W_{2}(\bz|\mathcal{C}_{i})  \overset{\bigtriangleup}{=} 
\frac {1}{M_{y}} \sum_{\bx \in \mathcal{C}_{i}} 
W_{2}(\bz|\bx) = \frac {1}{M_{y}} \sum_{j=0}^{M_{y}-1} W_{2}(\bz|\bx_{i,j}).
\end{equation}

Let $\mathbf{Y} \in \mathcal{Y}^{n}$ and $\mathbf{Z} \in \mathcal{Z}^{n}$ be the channel outputs resulting from the transmission of $\mathbf{X}_{i,j}$.
Define the average error probabilities of decoders (\ref{MLstrong}) and
(\ref{MLweak}) as
\begin{align}
\bar{P}_{\mbox{\tiny e,s}}(R_{y}, R_{z}) &=  \frac {1}{M_{y} M_{z} } \sum_{i=0}^{M_{z}-1} \sum_{j=0}^{M_{y}-1} \mathrm{Pr} 
\Big\{ [\hat{i}(\mathbf{Y}), \hat{j}(\mathbf{Y})] \neq (i,j) \Big| \mathbf{X}_{i,j}~\mbox{sent} \Big\} ,
\end{align}
and
\begin{align}
\bar{P}_{\mbox{\tiny e,w}} (R_{y}, R_{z})
&=  \frac {1}{M_{y} M_{z} } \sum_{i=0}^{M_{z}-1} \sum_{j=0}^{M_{y}-1}  \mathrm{Pr} \Big\{ \tilde{i}(\mathbf{Z}) \neq i  \Big| \mathbf{X}_{i,j}~\mbox{sent}  \Big\} ,
\end{align}
where in both definitions, $\mathrm{Pr} \{ \cdot \}$ 
designates probabilities associated with the 
randomness of the codebook, as well as that of the channel outputs given its input.
The corresponding random coding error exponents are defined as
\begin{equation}
\label{StrongExponent}
 E_{\mbox{\tiny s}}(R_{y}, R_{z}) = \lim_{n \to \infty} \Bigg[ - \frac {\ln  \bar{P}_{\mbox{\tiny e,s}} (R_{y}, R_{z})}{n}  \Bigg],
\end{equation} 
and
\begin{equation}
\label{WeakExponent}
 E_{\mbox{\tiny w}}(R_{y}, R_{z}) = \lim_{n \to \infty} \Bigg[ - \frac {\ln  \bar{P}_{\mbox{\tiny e,w}} (R_{y}, R_{z})}{n}  \Bigg],
\end{equation}
provided that the limits exist. Our main objective is to obtain 
single--letter expressions for 
$E_{\mbox{\tiny s}}(R_{y}, R_{z})$ and $E_{\mbox{\tiny w}}(R_{y},
R_{z})$.
As for the universal decoders, consider first the weak user. 
We wish to find a function $F(\bz,\bu_{i},\mathcal{C}_{i})$, 
that is independent of the (unknown) parameters of the channel $W_2$,
such that the following
{\it universal decoder} for the weak user
\begin{equation}
\label{GeneralWeak}
  \tilde{i}_{ \mbox{\tiny U} }(\bz) = \operatorname*{arg\,max}_{0 \leq i \leq M_{z}-1} F(\bz,\bu_{i},\mathcal{C}_{i}) 
\end{equation}
achieves an average error probability whose exponent is $E_{\mbox{\tiny
w}}(R_{y}, R_{z})$.
By the same token, we wish to find a universal decoder for the strong
user, of the form 
\begin{equation}
\label{GeneralStrong}
[\hat{i}_{ \mbox{\tiny U} }(\by), \hat{j}_{ \mbox{\tiny U} }(\by)] 
= \operatorname*{arg\,max}_{0 \leq i \leq M_{z}-1, 0 \leq j \leq M_{y}-1} G(\by,\bu_{i},\bx_{i,j}) ,
\end{equation}
where the function $G$ is independent of $W_1$, yet the decoder
$[\hat{i}_{ \mbox{\tiny U} }(\by), \hat{j}_{ \mbox{\tiny U}
}(\by)]$ achieves $E_{\mbox{\tiny s}}(R_{y}, R_{z})$.

\section {Main Results}

\subsection {Exact Random Coding Error Exponents}
Let $Q_{UXY}$ and $Q_{UXZ}$ denote two generic joint probability distributions of 
the random vectors $(U,X,Y)$ and $(U,X,Z)$, 
whose $(UX)$-marginals are both
identical to $P_{UX}$. 
Define
\begin{align}
\label{InnerFunctionStrong}
  E_{y}\big(Q_ {UXY}, R_{y}, R_{z}\big) = 
\min  \Big\{  \big[  I_{Q}(U;Y)  +  &  \big[ I_{Q}(X;Y|U) - R_{y} \big]_{+} - R_{z}  \big]_{+}, \nonumber \\ &\big[ I_{Q}(X;Y|U) - R_{y} \big]_{+}    \Big\},
\end{align}
and
\begin{equation}
\label{InnerFunctionWeak}
  E_{z}\big(Q_ {UXZ}, R_{y}, R_{z}\big) = 
 \big[  I_{Q}(U;Z)  +  \big[ I_{Q}(X;Z|U) - R_{y} \big]_{+} - R_{z}  \big]_{+}.
\end{equation}
Our first main result is the following. \\
\textbf{Theorem 1.} Under the assumptions of Section 2, the limits
(\ref{StrongExponent}) and (\ref{WeakExponent}) exist and are given by the following single--letter expressions: 
\begin{align}
\label{StrongExponents}
E_{ \mbox{\tiny s} }(R_{y}, R_{z}) &=
\min_{ Q_{Y|UX}} \Big\{  D(Q_{Y|UX} \| W_{Y|X} |  P_{UX}   )  +  E_{y}\big(Q_ {UXY}, R_{y}, R_{z}\big)   \Big\},  \\
\label{WeakExponents}
E_{ \mbox{\tiny w} }(R_{y}, R_{z}) &=
\min_{ Q_{Z|UX}} \Big\{  D(Q_{Z|UX} \| W_{Z|X} |  P_{UX}   )  +  E_{z}\big(Q_ {UXZ}, R_{y}, R_{z}\big)   \Big\}.
\end{align}
We prove the result concerning the strong user in Section 4 and the result concerning the weak user in Sections 5.
Notice that both error exponents depend on both coding rates, 
in contrast to the error exponents given 
in the previous works \cite{GALLAGER74} and \cite{KS80}. 

Several remarks are now in order. \\
$\bullet$ An immediate byproduct of Theorem 1 is finding the set of rate pairs $(R_{y}, R_{z})$ for which both $E_{ \mbox{\tiny s} }(R_{y}, R_{z}) >0$ and $E_{ \mbox{\tiny w} }(R_{y}, R_{z}) >0$. It is not difficult to show that this set is given by:
\begin{equation}
  \mathcal{R} = \big\{(R_{y}, R_{z}) |~ R_{y}< I(X;Y|U),~R_{y}+R_{z}< I(X;Y),~R_{z}< I(U;Z)  \big\},
\end{equation}
evaluated with the distribution $P_{UX} \times W_{Y|X} \times W_{Z|X}$. The convex hull of the closure of the union over all code distributions $\{ P_{UX} \}$ gives the capacity region. We may also consider an {\it individual attainable region} for each user, i.e., the set of rate pairs for which the probability of error vanishes for one of the users, but without taking into account the other user. Later on, individual attainable regions will become relevant when we consider the phase diagrams. It is not difficult to show that the attainable region for the weak user, to be denoted by $\mathcal{R}_{ \mbox{\tiny w} }$, is given by $   \mathcal{R}_{ \mbox{\tiny w} } = \big\{(R_{y}, R_{z}) |~ R_{y}+R_{z}< I(X;Z)  \big\} \cup   \big\{(R_{y}, R_{z}) |~R_{z}< I(U;Z)  \big\}$,
evaluated with the distribution $P_{UX} \times W_{Z|X}$, while the attainable region for the strong user, to be denoted by $\mathcal{R}_{ \mbox{\tiny s} }$, is given by
$  \mathcal{R}_{ \mbox{\tiny s} } = \big\{(R_{y}, R_{z}) |~ R_{y}+R_{z}< I(X;Y)  \big\} \cap   \big\{(R_{y}, R_{z}) |~R_{y}< I(X;Y|U)  \big\}$,
evaluated with the distribution $P_{UX} \times W_{Y|X}$. Notice that the attainable region of the weak user is not bounded, i.e., reliable bin index decoding may still be guaranteed for any satellites rate $R_{y}$, as long as $R_{z}< I(U;Z)$.
\\
$\bullet$ The computation of the error exponents involves minimizations over auxiliary channels $Q_{Y|UX}$ and $Q_{Z|UX}$. For large input and output alphabets, we are motivated to look for alternative expressions for the error exponents, whose optimization does not depend on the alphabet sizes, even at the expense of some loss in the exponential tightness. We will discuss such an alternative form in the sequel. 
\\
$\bullet$ Both error exponents depend on the input distribution. While in the single-user regime, we may maximize the final expression over the input distribution in order to maximize the error exponent, this is no longer the case for the ABC. Even in the simplest case of a binary symmetric BC, we see that the best code for the strong user is the worst one for the weak user, and vice versa. To see why is that true, let $P_{U}=(\frac{1}{2}, \frac{1}{2})$ and let $P_{X|U}$ be a BSC with a crossover probability $\frac{1}{2}$. In this case, the hierarchy of the codebook degenerates, i.e., the codebook has a constant composition, which is best for the strong user. In the other extreme, $P_{X|U}$ is a BSC with a crossover probability $0$. The error probability of the strong user is almost one, but the error exponent of the weak user is the largest and independent of $R_{y}$.   
Hence, the choice of the input distribution trades off between the error exponents of the two users.
\\
$\bullet$ As can be seen from the minimum in eq. (\ref{InnerFunctionStrong}), there are two different kinds of error events for the strong user. Let $Q^{*}$ denote the minimizer in (\ref{StrongExponents}). Now, if for some $(R_{y}, R_{z})$, the inequality $\big[  I_{Q^{*}}(U;Y)  +  \big[ I_{Q^{*}}(X;Y|U) - R_{y} \big]_{+} - R_{z}  \big]_{+} > \big[ I_{Q^{*}}(X;Y|U) - R_{y} \big]_{+}$ holds, then the dominant error event for the strong user is caused by competing codewords from the true cloud, otherwise, the dominant error event is caused by competitive clouds. 
\\
$\bullet$ In fact, the cardinality $|\calU|$ is a free parameter in our problem. As such, we may let $|\calU| \to \infty$, and it is definitely not obvious that a finite $|\calU|$ is optimal. This is because we cannot see how to apply the usual cardinality bounding techniques based on the support lemma \cite[page 310]{CK11}. It must be clear that even if the optimal $|\calU|$ is finite, it may not be the same as the bound given in the converse theorem of the capacity region of the ABC ($|\calU| \leq |\calX| + 2$) \cite{KM77}.

\subsection {Universal Decoders}

As mentioned in the Introduction, universal MMI decoders for both 
receivers were proposed in \cite{KS80}, where for the weak user, this decoder was defined by:  
\begin{equation}
\label{KS_weak}
  \tilde{i}_{ \mbox{\tiny KS} }(\bz) = \operatorname*{arg\,max}_{0 \leq i \leq M_{z}-1}  \hat{I}_{\bu_{i}\bz}(U;Z) . 
\end{equation}
The error exponent of such a decoder is 
inferior to the error exponent of the optimal (ML) decoder, 
because for one thing, it makes no use of $\{\calC_i\}$, but only of the cloud centers. 
The universal decoder (\ref{KS_weak}) achieves the following error exponent \cite{KS80}
\begin{align}
E_{ \mbox{\tiny w,KS} }(R_{z}) =
\min_{ Q_{Z|UX}}  \Big\{  &D(Q_{Z|UX} \| W_{Z|X} |  P_{UX}   ) 
+ [ I_{Q}(U;Z) - R_{z} ]_{+}   \Big\},
\end{align}
and by comparing it numerically to (\ref{WeakExponents}) in the case of the binary symmetric BC (see Subsection 3.4), 
it is evident that $E_{\mbox{\tiny w}}(R_{y}, R_{z})$ can be strictly higher than $E_{ \mbox{\tiny w,KS}}(R_{z})$, 
due to the additional term in (\ref{InnerFunctionWeak}).  
Hence, one may wonder whether a different universal decoder exists, 
whose error exponent is as large as $E_{\mbox{\tiny w}}(R_y, R_{z})$.
It turns out that the answer to this question is affirmative, and indeed,
this universal decoder relies entirely on $\calC$ and $\{\bu_i\}$.
In Section 5, we prove the following theorem. \\
\textbf{Theorem 2.} Define the function
\begin{align}
\label{WeakFunction}
F(\bz, \bu_{i}, \mathcal{C}_{i}) 
= \max_{0 \leq j \leq M_{y}-1} \Big\{ \hat{I}_{\bu_{i}\bz}(U;Z)  +  [ \hat{I}_{\bu_{i}\bx_{ij}\bz}(X;Z|U)  - R_{y} ]_{+} \Big\} .
\end{align}
The universal decoder
\begin{align}
  \tilde{i}_{ \mbox{\tiny U} }(\bz) = \operatorname*{arg\,max}_{0 \leq i \leq M_{z}-1}   F(\bz, \bu_{i}, \mathcal{C}_{i})   
\end{align}
achieves $E_{ \mbox{\tiny w} }(R_{y}, R_{z})$. 

It turns out that there is also another universal decoder (with the same error
exponent), whose structure is much more similar
to the ML decoder of (\ref{MLweak}), 
in the sense that its metric is based on  
summation over $\calC_i$, except that here, the unknown likelihood
function is replaced by the exponentiated empirical mutual information.
In the Appendix we prove the following theorem. \\
\textbf{Theorem 3.} The universal decoder
\begin{equation}
  \tilde{i}_{ \mbox{\tiny U} }(\bz) = \operatorname*{arg\,max}_{0 \leq i \leq M_{z}-1} 
  \left\{    \sum_{j=0}^{M_{y}-1} 
e^{n   \hat{I}_{\bu_{i}\bx_{ij}\bz}(UX;Z)   }  \right\}  
\end{equation}
achieves $E_{ \mbox{\tiny w} }(R_{y}, R_{z})$. 

We next proceed to the strong user and present a universal decoder.
It turns out that the MMI--like metric 
of the universal bin index decoder, as given in Theorem 2 (but with $\bz$
replaced by $\by$), works well also for the strong user. 
The main difference between them is rooted in the way they use the metric. 
While the weak user first maximizes it within each cloud, 
and only then finds the cloud with the maximal value, 
the strong user maximizes it over both indices simultaneously. 
More precisely, we claim the following, which is proved in Section 4. \\
\textbf{Theorem 4.} Define the function
\begin{align}
\label{StrongFunction}
G(\by, \bu_{i}, \bx_{ij}) = \hat{I}_{\bu_{i}\by}(U;Y)  +  [ \hat{I}_{\bu_{i}\bx_{ij}\by}(X;Y|U)  - R_{y} ]_{+}   . 
\end{align}
The universal decoder
\begin{align}
  [\tilde{i}_{ \mbox{\tiny U} }(\by), \tilde{j}_{ \mbox{\tiny U}
}(\by)]
= \operatorname*{arg\,max}_{0 \leq i \leq M_{z}-1, 0 \leq j \leq M_{y}-1}  
  G(\by, \bu_{i}, \bx_{ij})  
\end{align}
achieves $E_{ \mbox{\tiny s} }(R_{y}, R_{z})$. 

At this point, it is interesting to compare
$[\tilde{i}_{ \mbox{\tiny U} }(\by), \tilde{j}_{ \mbox{\tiny U}
}(\by)]$
to the universal decoder of the strong user in \cite{KS80},  
\begin{equation}
\label{KS_Strong}
[\hat{i}_{ \mbox{\tiny KS} }(\by), \hat{j}_{ \mbox{\tiny KS} }(\by) ] = \operatorname*{arg\,max}_{0 \leq i \leq M_{z}-1, 0 \leq j \leq M_{y}-1}  \hat{I}_{\bu_{i}\bx_{ij}\by}(UX;Y),
\end{equation}
and whose random coding error exponent is given by \cite{KS80}
\begin{equation}
E_{ \mbox{\tiny s,KS} }(R_{y}, R_{z}) =
\min_{ Q_{Y|UX}} \Big\{  D(Q_{Y|UX} \| W_{Y|X} |  P_{UX}   )  +  E_{y,\mbox{\tiny KS}}\big(Q_ {UXY}, R_{y}, R_{z}\big)   \Big\},
\end{equation}
where 
\begin{equation}
\label{KSinnerStrong}
  E_{y,\mbox{\tiny KS}}\big(Q_ {UXY}, R_{y}, R_{z}\big) = 
\min  \Big\{  \big[  I_{Q}(UX;Y)  - (R_{y} + R_{z})  \big]_{+}, \big[ I_{Q}(X;Y|U) - R_{y} \big]_{+}    \Big\}.
\end{equation}
By the identity $I(UX;Y) = I(U;Y) + I(X;Y|U)$, it is 
easy to see that $E_{ \mbox{\tiny s,KS} }(R_{y}, R_{z}) = E_{
\mbox{\tiny s}}(R_{y}, R_{z})$, proving that (\ref{KS_Strong}) has an error exponent as that of (\ref{MLstrong}), a fact that was not asserted in \cite{KS80}.

\subsection {Gallager--Style Lower Bounds}

As mentioned before, the calculations of (\ref{StrongExponents}) and (\ref{WeakExponents}) 
involve minimizations over auxiliary channels, 
which become painful when the input and output alphabets are large. 
For this reason, we look for other forms of error exponent formulas, 
where the number of parameters to be optimized does not grow with the alphabet sizes, 
but the price of this might be some loss in the tightness of the bounds, 
i.e., we obtain {\it lower bounds} on the random coding error exponents. Even in the single user case, the random coding error exponent involves a minimization over an auxiliary channel, where Csisz\'{a}r and K\"{o}rner \cite[Exercise 10.24]{CK11} show that the exact error exponent is lower bounded by the following expression 
\begin{equation}
\label{SingleUserGallager}
E_{ \mbox{\tiny G} }(R) =
\max_{ \rho \in [0,1] } \Bigg\{  - \log \sum_{y} \bigg[ \sum_{x} P(x)W^{\frac{1}{1+\rho}}(y|x)     \bigg]^{1+\rho}  -\rho R   \Bigg\},
\end{equation}
where the subscript 'G' stands for ``Gallager", who was the first to derive and analyze the error exponent in this form \cite{GAL68}. It is important to note that for the optimal code distribution, (\ref{SingleUserGallager}) is not only a lower bound, but the exact random coding error exponent \cite{CK11}. It turns out that the exact random coding error exponents of the two users in the ABC can be lower bounded by the same methods as in \cite{CK11}. In Section 6, we prove the following theorem. \\
\textbf{Theorem 5.} Define the functions
\begin{align}
\Phi(u,y,s) &= \sum_{x} P(x|u)[W_{1}(y|x)]^{\frac{1}{1+s}}, \\
\Psi(u,z,s) &= \sum_{x} P(x|u)[W_{2}(z|x)]^{\frac{1}{1+s}}. 
\end{align}
The exact random coding error exponent of the strong user is lower bounded by
\begin{align}
E_{ \mbox{\tiny s} }(R_{y}, R_{z}) \geq
\min \Big\{  E_{ y,1 }(R_{y}), E_{ y,2 }(R_{y}, R_{z})      \Big\}, 
\end{align}
where
\begin{align}
\label{StrongTerm1}
E_{ y,1 }(R_{y}) &= \max_{\rho \in [0,1] } \Bigg\{  - \sum_{u} P(u) \log \Bigg( \sum_{y} 
 \Phi^{1+\rho}(u,y,\rho)  \Bigg)  - \rho R_{y}  
  \Bigg\},   \\
\label{StrongTerm2}
E_{ y,2 }(R_{y}, R_{z}) &= \max_{\mu \in [0,1] } \max_{\lambda \in [0,\mu] } \Bigg\{    
  -  \log \Bigg[ \sum_{y}  \Bigg(  \sum_{u} P(u)  \Phi^{\frac{1+\lambda}{1+\mu}}(u,y,\lambda)    \Bigg)^{1+\mu}  \Bigg]   -  \lambda R_{y}  -\mu R_{z}  \Bigg\}  .
\end{align}
In addition, the random coding error exponent 
of the weak user is lower bounded by
\begin{align}
\label{LowerBoundWeak}
&E_{ \mbox{\tiny w} }(R_{y}, R_{z}) \geq
\max_{\mu \in [0,1] } \max_{\lambda \in [0,\mu] } \Bigg\{    
  -  \log \Bigg[ \sum_{z}  \Bigg(  \sum_{u} P(u)  \Psi^{\frac{1+\lambda}{1+\mu}}(u,z,\lambda)    \Bigg)^{1+\mu}  \Bigg]   -  \lambda R_{y}  -\mu R_{z}  \Bigg\}  .
\end{align}
\\
These lower bounds 
involve maximizations over one or two parameters only, in contrast to the original error exponents, and so, they are much easier to evaluate. In Section 7, we study them and show how they behave in different regions of the plane of rates. In contrast to the single user case, both lower bounds of the two users depend on the code distribution, but now we are no longer able to optimize both of them simultaneously, for the reason we mentioned above, in subsection 3.1.

\subsection {Numerical Results and Phase Diagrams}
We next provide some numerical results, comparing our exponents to those of \cite{KS80} and \cite{KM11}. 
Let $W_1$ and $W_2$ be two binary symmetric channels (BSC´s) with
crossover parameters $p_y$ and $p_z$, respectively ($p_z > p_y$).
Let $\mathcal{U}$ be binary as well and let $P_{U}$ be 
uniformly distributed over $\{0,1\}$. 
Also, let $P_{X|U}$ be a BSC with crossover parameter $\beta \in [0,1]$.  \\
The capacity region of our model is given by:
\begin{align}  
R_{z} &\leq \ln 2 - h(\beta * p_{z})  \nonumber  \\
R_{y} &\leq h(\beta * p_{y}) - h( p_{y}), 
\end{align}
where $\beta * p = \beta(1-p) + (1 - \beta)p$ and $h(x)$ is the binary entropy function.

\subsubsection{Gallager-Style Lower Bounds} 
Using Theorem 5, we find that for the strong user,
\begin{align}
\label{StrongLowerBound}
E_{ \mbox{\tiny s} }(R_{y}, R_{z}) &\geq
\min \Big\{ E_{y,1}(R_{y}),  E_{y,2}(R_{y}, R_{z})    \Big\},
\end{align}
where,
\begin{align}
&E_{y,1}(R_{y}) =
\max_{\rho \in [0,1] } \Bigg\{  -  \log \Bigg\{  \bigg[  (1-\beta)(1-p_{y})^{\frac{1}{1+\rho}} + \beta \cdot p_{y}^{\frac{1}{1+\rho}}     \bigg]^{1+\rho}  \nonumber \\ &\;\;\;\;\;\;\;\;\;\;\;\;\;\;\;\;\;\;\;\;\;\;\;\;\;\;\;\;\;\;\;\;\;\;\;\;\;\;\;\;\;\;\;\;\;  +  \bigg[  (1-\beta) \cdot p_{y}^{\frac{1}{1+\rho}} + \beta \cdot (1-p_{y})^{\frac{1}{1+\rho}}     \bigg]^{1+\rho}   \Bigg\}   - \rho R_{y}    \Bigg\}, \\
&E_{y,2}(R_{y}, R_{z}) \nonumber \\
&= \max_{\mu \in [0,1] } \max_{\lambda \in [0,\mu] } \Bigg\{  -  \ln 2  - (1+\mu) \cdot \log  \Bigg\{  \frac{1}{2} \cdot \bigg[ (1-\beta)(1-p_{y})^{\frac{1}{1+ \lambda}} + \beta \cdot p_{y}^{\frac{1}{1+\lambda}}    \bigg]^{\frac{1+\lambda}{1+\mu}} \nonumber \\ & \;\;\;\;\;\;\;\;\;\;\;\;\;\;\;\;\;\;\;\;\;\;\;\;\;\;\;\;\; +  \frac{1}{2} \cdot \bigg[  (1-\beta) \cdot p_{y}^{\frac{1}{1+ \lambda}} + \beta \cdot (1-p_{y})^{\frac{1}{1+ \lambda}}    \bigg]^{\frac{1+\lambda}{1+\mu}}  \Bigg\}    - \lambda R_{y}  -\mu R_{z}    \Bigg\}.
\end{align}
For the weak user,
\begin{align}
\label{WeakLowerBound}
&E_{ \mbox{\tiny w} }(R_{y}, R_{z}) \nonumber \\
&\geq \max_{\mu \in [0,1] } \max_{\lambda \in [0,\mu] } \Bigg\{  -  \ln 2  - (1+\mu) \cdot \log  \Bigg\{  \frac{1}{2} \cdot \bigg[ (1-\beta)(1-p_{z})^{\frac{1}{1+ \lambda}} + \beta \cdot p_{z}^{\frac{1}{1+\lambda}}    \bigg]^{\frac{1+\lambda}{1+\mu}} \nonumber \\ & \;\;\;\;\;\;\;\;\;\;\;\;\;\;\;\;\;\;\;\;\;\;\;\;\;\;\;\;\; +  \frac{1}{2} \cdot \bigg[  (1-\beta) \cdot p_{z}^{\frac{1}{1+ \lambda}} + \beta \cdot (1-p_{z})^{\frac{1}{1+ \lambda}}    \bigg]^{\frac{1+\lambda}{1+\mu}}  \Bigg\}    - \lambda R_{y}  -\mu R_{z}    \Bigg\}.
\end{align}
We present the lower bounds by plotting families of curves, one for each exponent, as a function of one rate, while the other rate is kept fixed. 
Let us choose the channel probabilities to be $p_{y} = 0.05$ and $p_{z} = 0.1$, and $\beta = 0.25$. 
In Fig.\ 1, we plot lower bounds to $E_{\mbox{\tiny s}}(R_{y}, R_{z})$ as a function of $R_{y}$, as given by (\ref{StrongLowerBound}), where $R_{z}$ takes five different values.
\begin{figure}[ht!]
\centering
\includegraphics[width=110mm]{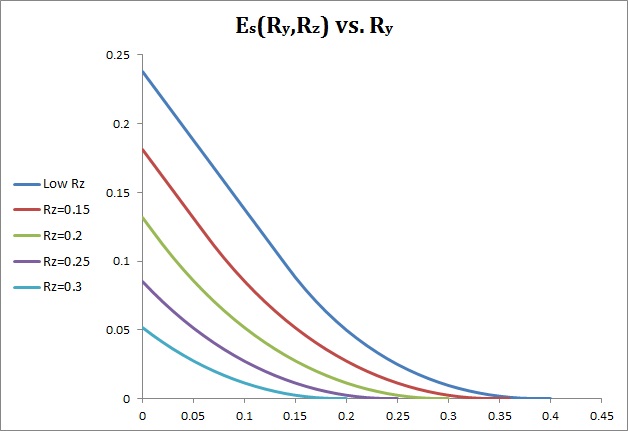}
\caption{Error exponents for the strong user for different values of $R_{z}$. \label{overflow}}
\end{figure}
As long as $R_{z} < 0.09$, the dominant error event is caused by wrong codewords from the true cloud. In this case, the error exponent is independent of the number of clouds and is given by the dark blue curve. As $R_{z}$ increases more, we find that above some critical rate, the error exponent begins to depend on the number of clouds, since the dominant error event is due to wrong codewords from competitive clouds. When the rate of the weak user is high, i.e., when the exponential number of clouds is higher than the capacity of the channel to the strong user ($R_{z} > 0.49 \approx h(0.05)$), reliable communication is no longer possible. 

In Fig.\ 2, we plot lower bounds to $E_{\mbox{\tiny w}}(R_{y}, R_{z})$ as a function of $R_{z}$, as given by (\ref{WeakLowerBound}), where $R_{y}$ takes five different values.
\begin{figure}[ht!]
\centering
\includegraphics[width=110mm]{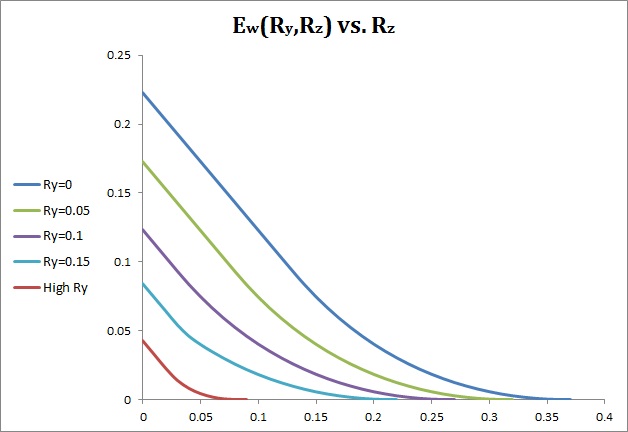}
\caption{Error exponents for the weak user for different values of $R_{y}$. \label{overflow}}
\end{figure}
At $R_{y} = 0$, we should obtain the error exponent of a single user. In this case, the numerical value at zero-rate is given by $E_{\mbox{\tiny w}}(0,0) = 0.22314$, and $E_{\mbox{\tiny w}}(0,R_{z})$ vanishes at $R_{z} \cong 0.36 \approx h(0.1)$, which is the capacity of the channel to the weak user. For $R_{y} > 0.32$, $E_{\mbox{\tiny w}}(R_{y}, R_{z})$ becomes independent of $R_{y}$, and is given by the red curve. In this case, we get a lower bound to the error exponent of the equivalent binary symmetric channel from the cloud center $U$ to the channel output of the weak user $Z$.

\subsubsection{Exact Exponents}
As for the exact random coding error exponents, given by Theorem 1, the optimization problems require the minimization over the auxiliary channels $Q_{Y|UX}$ and $Q_{Z|UX}$. Let us compare the Gallager-style lower bounds to the exact exponents. In Fig.\ 3, we see two pairs of curves of the exact exponents and their lower bounds, where $R_{z}=0.05$ and $\beta$ takes two different values. The exact exponents are strictly better than the Gallager-style exponents. Similar results are obtained for the weak user as well (not shown here). It is important to note that in some regions in the $R_{y}-R_{z}$ plane, the lower bounds are equal to the exact random coding error exponents.
\begin{figure}[ht!]
\centering
\includegraphics[width=110mm]{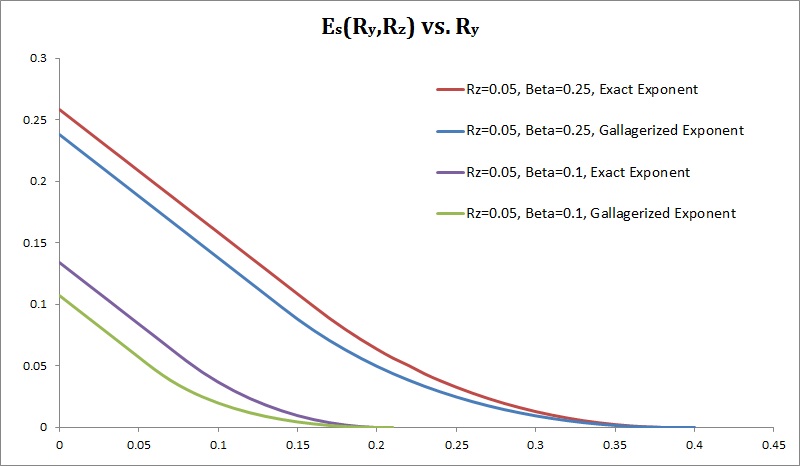}
\caption{Exact exponents and lower bound for the strong user. \label{overflow}}
\end{figure}

\subsubsection{Comparison with Previous Works}
As far as we know, no other works on universal decoding for the ABC exists, besides the one of \cite{KS80}. Although the error exponent of the strong user given there is optimal w.r.t.\ the ML decoder, it is not the case for the weak user. The universal decoder of \cite{KS80} for the weak user uses only the cloud centers and is independent of $R_{y}$, while the new universal decoder of Theorem 2 makes use of the entire codebook, which is the main reason for the resulted improvement. The difference between the error exponents is larger for lower values of $R_{y}$. 
As before, let $p_{z} = 0.1$ and $\beta = 0.25$. 
Fig.\ 4 demonstrates the difference between the error exponents of the two universal decoders in the extreme case of $R_{y} = 0$.
\begin{figure}[ht!]
\centering
\includegraphics[width=110mm]{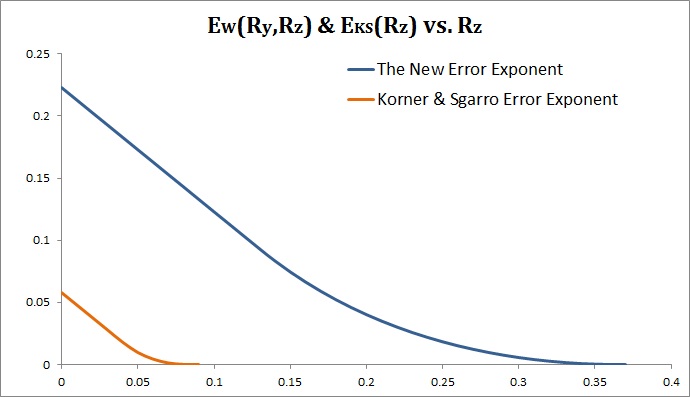}
\caption{Comparison to \cite{KS80} for the weak user. \label{overflow}}
\end{figure}

To the best of our knowledge, the most up-to-date work on exponential lower bounds to the reliability functions of the ABC is \cite{KM11}, where random coding error exponents were derived using two different techniques. Each of those derivations includes at least one step that may not be exponentially tight.
 Also, in \cite{KM11}, the random codebooks are assumed to be drawn i.i.d.. We expect our proposed exact random coding error exponents to improve on \cite{KM11}, because of two reasons: first, our analysis is exponentially tight, and second, our ensemble is of the uniform distribution across types. This kind of random codes are known \cite[Section 7.3]{GAL68} to be better than the i.i.d.\ ensembles. Our comparison here focuses on the error exponent of the weak user only. Again, let $p_{z} = 0.1$, $\beta = 0.25$ and $R_{y} = 0.4$. 
Fig.\ 5 compares the two error exponents, and shows that the new exponent is better.  
\begin{figure}[ht!]
\centering
\includegraphics[width=110mm]{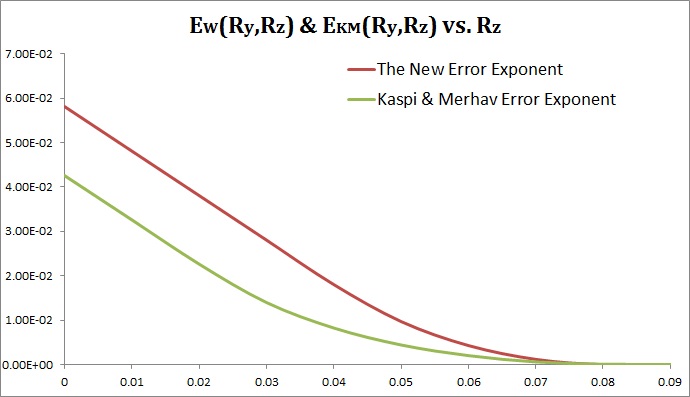}
\caption{Comparison to \cite{KM11} for the weak user. \label{overflow}}
\end{figure}

\subsubsection{Phase Diagrams}
In the single user case, it is known that the error exponent behaves differently in different ranges of rates, i.e., it is affine at low rates and curvy at high rates. By the same token, for the ABC, the plane of rates can be divided into several different regions, where in each one of them, the error exponents behaves differently. This partition of the plane of rate pairs is of course, more involved than in the single-user case. We refer to it as a phase diagram, a term borrowed from physics.
In order to study the various types of behavior of the lower bound of Theorem 5, let us invoke the following alternative and equivalent lower bound for the random coding error exponent of the weak user
\begin{equation}
\label{AltLowerbound}
E_{ \mbox{\tiny w} }(R_{y}, R_{z}) \geq
\max_{\mu \in [0,1] } \max_{s \in [0,1] } \Bigg\{    
  -  \log \Bigg[ \sum_{z}  \Bigg(  \sum_{u} P(u)  \Psi^{\frac{1+s \mu}{1+\mu}}(u,z, s\mu)    \Bigg)^{1+\mu}  \Bigg]   -  s \mu R_{y}  -\mu R_{z}  \Bigg\}  .
\end{equation}
Since the maximization region is now the unit square, this form is more convenient to analyze than that of (\ref{LowerBoundWeak}). 
Fig.\ 6 displays a partition of the plane $R_{y}-R_{z}$ to different regions for the Gallager-style lower bound of the weak user, where $\beta=0.1$, and $p_{z} = 0.1$. Although not shown here, the phase diagrams of the exact exponents behave similarly.
\begin{figure}[ht!]
\centering
\includegraphics[width=140mm]{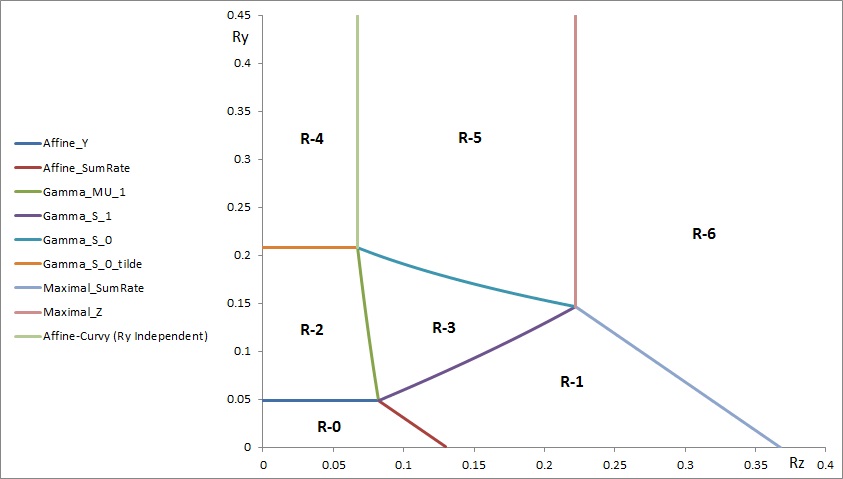}
\caption{Phase diagram for the weak user ($\beta=0.1$). \label{overflow}}
\end{figure}
The study in Section 7 provides a characterization of the different regions 
from the viewpoint of the type of dependence of the error exponent upon
the rates and the maximizers $s^{*}$ and $\mu^{*}$ (see Table 1).

\begin {table}
\begin{center}
  \begin{tabular}{ | c | c | c | c |}
    \hline
    Region & Dependence on $(R_{y},R_{z})$ & $\mu^{*}$ & $s^{*}$ \\ \hline
    \textbf{R--0} & affine in $R_{y}+R_{z}$ & 1 & 1 \\ \hline
    \textbf{R--1} & curvy in $R_{y}+R_{z}$ & [0,1) & 1 \\ \hline
    \textbf{R--2} & curvy in $R_{y}$, affine in $R_{z}$ & 1 & [0,1) \\ \hline
    \textbf{R--3} & curvy in both $R_{y}$ and $R_{z}$ & [0,1) & [0,1) \\ \hline
    \textbf{R--4} & affine in $R_{z}$, independent of $R_{y}$ & 1 & 0 \\ \hline
    \textbf{R--5} & curvy in $R_{z}$, independent of $R_{y}$ & [0,1) & 0 \\ \hline
    \textbf{R--6} & vanishes for all $(R_{z}, R_{y})$ & 0 & 0 \\ 
    \hline
  \end{tabular}
\caption {Dependence of (\ref{AltLowerbound}) on $R_{y}$ and $R_{z}$ in various regions in the plane (see Fig.\ 6).}
\end{center}
\end {table}

\section {Universal decoding for the strong user}

\subsection {Analysis for a General Decoder}

Let us first derive the exact random coding error exponent for a strong user that uses the following generic decoder
\begin{align}
\label{genericStrong}  
  [\hat{i} (\by), \hat{j} (\by)] =   \operatorname*{arg\,max}_{0 \leq i \leq M_{z}-1, 0 \leq j \leq M_{y}-1}   f(Q_{U_{i}X_{ij}Y}),    
\end{align}
where from now on, $Q_{UXY}$ will designates the joint empirical distribution induced by the three sequences $\bu$, $\bx$ and $\by$, i.e., $Q_{UXY} = \hat{P}_{\bu \bx \by}$. 
The average error probability $\bar{P}_{e}(R_{y},R_{z},n)$, associated with (\ref{genericStrong}) is 
\begin{equation}
\begin{aligned}
 \bar{P}_{e}(R_{y},R_{z},n)      \overset{\bigtriangleup}{=}  \frac{1}{M_{y} M_{z}} \sum_{i=0}^{M_{z}-1} \sum_{j=0}^{M_{y}-1} &\mathrm{Pr} \Bigg\{   \bigg\{ \bigcup_{k \neq j}\Big\{ f(Q_{U_{i}X_{ik}Y}) \geq f(Q_{U_{i}X_{ij}Y}) \Big| \mathbf{X}_{ij} \; \mathrm{sent}    \Big\}    \bigg\} \\     &\bigcup      \bigg\{ \bigcup_{l \neq i} \bigcup_{k} \Big\{ f(Q_{U_{l}X_{lk}Y}) \geq f(Q_{U_{i}X_{ij}Y}) \Big| \mathbf{X}_{ij} \; \mathrm{sent}   \Big\}   \bigg\}       \Bigg\} 
\end{aligned}
\end{equation}
where $\mathrm{Pr}\{\cdot\}$ pertains to the randomness of the codebook as well as that of the channel output given its input. Without loss of generality, we assume throughout, that the transmitted codeword is $\mathbf{X}_{00} = \bx_{00}$. 
We define
\begin{equation}
   \mathcal{A} \overset{\bigtriangleup}{=} \bigcup_{k=1}^{M_{y}-1} \mathcal{A}_{k}    \overset{\bigtriangleup}{=} \bigcup_{k=1}^{M_{y}-1}\big\{  f(Q_{U_{0}X_{0k}Y}) \geq f(Q_{U_{0}X_{00}Y})     \big\}
\end{equation}
and                                       
\begin{equation}
   \mathcal{B} \overset{\bigtriangleup}{=} \bigcup_{l=1}^{M_{z}-1} \mathcal{B}_{l}  \overset{\bigtriangleup}{=} \bigcup_{l=1}^{M_{z}-1} \bigcup_{k=0}^{M_{y}-1} \mathcal{B}_{lk}    \overset{\bigtriangleup}{=} \bigcup_{l=1}^{M_{z}-1} \bigcup_{k=0}^{M_{y}-1} \big\{ f(Q_{U_{l}X_{lk}Y}) \geq f(Q_{U_{0}X_{00}Y}) \big\} .
\end{equation}
Define the real number $s$ as 
\begin{equation}
  s  \overset{\bigtriangleup}{=}   f(Q_{U_{0}X_{00}Y})  .
\end{equation}
The pairwise average error probability, conditioned on the center of the competitive cloud, is given by
\begin{align}
\label{START1}
 \mathrm{Pr}(\mathcal{B}_{lk} | \mathbf{U}_{l} = \bu')   &\overset{\bigtriangleup}{=} 
	\mathrm{Pr} \Big\{ f(Q_{U_{l}X_{lk}Y}) \geq f(Q_{U_{0}X_{00}Y}) \Big| \mathbf{U}_{l} = \bu' \Big\} \\
\label{GUMP1a}
&=     \sum_{\{\bx': \;  f(Q_{U'X'Y})  \geq s  \}} P ( \bx' | \bu' ) \\
&=  \sum_{\{ Q_{X'|U'Y} \in \mathcal{S}(Q_{U'Y}) : ~  f(Q_{U'X'Y})  \geq s   \}} 
       \sum_{\btildex \in \mathcal{T}(Q_{X'|U'Y}|\bu',\by)}   P ( \btildex | \bu' ) \\
&=  \sum_{\{ Q_{X'|U'Y} \in \mathcal{S}(Q_{U'Y}) : ~  f(Q_{U'X'Y})  \geq s   \}} 
       P ( \bx' | \bu' ) \cdot  |\mathcal{T}(Q_{X'|U'Y}|\bu',\by)| \\
&\doteq  \sum_{\{ Q_{X'|U'Y} \in \mathcal{S}(Q_{U'Y}) : ~  f(Q_{U'X'Y})  \geq s   \}} \exp \Big\{-n \cdot  I_ {Q}(X;Y|U)   \Big\}  \\
\label{GUMP1b}
&\doteq  \max_{\{ Q_{X'|U'Y} \in \mathcal{S}(Q_{U'Y}) : ~  f(Q_{U'X'Y})  \geq s  \}} \exp \Big\{-n \cdot    I_ {Q}(X;Y|U)    \Big\}  \\
&=   \exp \Bigg\{-n \cdot  \min_{\{ Q_{X'|U'Y} \in \mathcal{S}(Q_{U'Y}) : ~  f(Q_{U'X'Y})  \geq s  \}}      I_ {Q}(X;Y|U)      \Bigg\}  \\
\label{END1}
&\overset{\bigtriangleup}{=}   \exp \Big\{-n \cdot   E_{0}(s, Q_{U'Y} )    \Big\}   ,
\end{align}
where $\mathcal{S}(Q_{UY})$ denotes the set of conditional distributions $\{Q_{X|UY}\}$ that are consistent with $P_{UX}$.
For a given $\mathbf{U}_{l} = \bu'$, the events $\{\mathcal{B}_{lk}\}_{k}$ are all pairwise independent since we have assumed that the various codewords are pairwise conditional independent given the cloud center. Using the exponential tightness of the truncated union bound \cite[Lemma A.2]{SHUL03}, we get
\begin{align}
 \mathrm{Pr} \Bigg\{  \bigcup_{k=0}^{M_{y}-1} \mathcal{B}_{lk} \Bigg| \mathbf{U}_{l} = \bu'   \Bigg\}   
&\doteq   \min  \Bigg\{ 1, \sum_{k=0}^{M_{y}-1} \mathrm{Pr}(\mathcal{B}_{lk} | \mathbf{U}_{l} = \bu')  \Bigg\}  \\  
&=   \min  \Big\{ 1, M_{y} \cdot  \mathrm{Pr}(\mathcal{B}_{l,0} | \mathbf{U}_{l} = \bu')  \Big\}  \\
&\doteq   \min  \bigg\{ 1, e^{nR_{y}} \cdot  \exp \Big[ -n \cdot   E_{0}(s, Q_{U'Y} )    \Big]  \bigg\} \\
&\overset{\bigtriangleup}{=}   \exp \Big\{-n \cdot   E_{1}(s, Q_{U'Y}  )    \Big\}  ,
\end{align}
where
\begin{equation}
  E_{1}(s, Q_{U'Y} ) =  \min_{Q_{X'|U'Y} \in \mathcal{S}(Q_{U'Y})} \Big\{\big[ I_ {Q}(X;Y|U) - R_{y} \big]_{+}: \:    f(Q_{U'X'Y})  \geq s  \Big\}.
\end{equation}
\\
Next, we obtain the probability of $\mathcal{B}_{l}$ by calculating the expectation w.r.t. the randomness of $\mathbf{U}_{l}$:
\begin{align}
\mathrm{Pr} \Big\{   \mathcal{B}_{l}   \Big\}  &=  \sum_{\bu' \in \mathcal{T}(P_{U})}  P_{U}(\bu')  \cdot  
 \mathrm{Pr} \Bigg\{  \bigcup_{k=0}^{M_{y}-1} \mathcal{B}_{lk} \Bigg| \mathbf{U}_{l} = \bu'   \Bigg\}   \\
\label{GUMP2a}
&\doteq \sum_{\bu' \in \mathcal{T}(P_{U})}    P_{U}(\bu')  \cdot  
  \exp \Big\{-n \cdot   E_{1}(s, Q_{U'Y} )    \Big\}   \\
&= \sum_{ \{ Q_{U'|Y} \in \mathcal{S}(Q_{Y})   \} }    \sum_{\btildeu \in \mathcal{T}(Q_{U'|Y}|\by)}    P_{U}(\btildeu)  \cdot  
  \exp \Big\{-n \cdot   E_{1}(s, Q_{\tilde{U}Y} )    \Big\}   \\
&= \sum_{ \{ Q_{U'|Y} \in \mathcal{S}(Q_{Y})   \} }     \frac{|\mathcal{T}(Q_{U'|Y}|\by)|}{|\mathcal{T}(\bu')|} \cdot    \exp \Big\{-n \cdot   E_{1}(s, Q_{U'Y} )    \Big\}   \\
&\doteq \sum_{ \{ Q_{U'|Y} \in \mathcal{S}(Q_{Y})   \} }      \exp \Bigg\{-n \cdot \bigg[  
I_ {Q}(U;Y)  +    E_{1}(s, Q_{U'Y} )    \bigg]  \Bigg\}      \\
\label{GUMP2b}
&\doteq   \exp \Bigg\{-n \cdot \min_{ \{ Q_{U'|Y} \in \mathcal{S}(Q_{Y})   \} } \bigg[  I_ {Q}(U;Y)  +   E_{1}(s, Q_{U'Y} )   \bigg]  \Bigg\} \\
&\overset{\bigtriangleup}{=}   \exp \Big\{-n \cdot   E_{2}(s, Q_{Y} )    \Big\},
\end{align}
where $\mathcal{S}(Q_{Y})$ is the set of all $\{Q_{U|Y}\}$ such that $\sum_{y}Q_{Y}(y)Q_{U|Y}(u|y)=P_{U}(u)$ for every $u \in \mathcal{U}$. 
Next, we turn to calculate the probabilities of the events $\mathcal{A}_{k}$. One can easily check that the entire derivation of eqs.\ (\ref{START1})-(\ref{END1}) holds in this case as well, except that now we condition on $\mathbf{U}_{0} = \bu_{0}$, such that the codewords are drawn from $P(\cdot | \bu_{0})$. We get 
\begin{align}
 \mathrm{Pr}(\mathcal{A}_{k} | \mathbf{U}_{0} = \bu_{0})   &\overset{\bigtriangleup}{=} 
	\mathrm{Pr} \Big\{ f(Q_{U_{0}X_{0k}Y}) \geq f(Q_{U_{0}X_{00}Y}) \Big| \mathbf{U}_{0} = \bu_{0} \Big\} \\
&\doteq  \exp \Big\{-n \cdot   E_{0}(s, Q_{U_{0}Y} )    \Big\} .
\end{align}
Notice that, for a given $\mathbf{U}_{0} = \bu_{0}$, $\mathbf{X}_{00} = \bx_{00}$ and $\mathbf{Y} = \by$, the events $\{\mathcal{A}_{k}\}$ (errors caused by codewords from the correct cloud) and $\{\mathcal{B}_{l}\}$ (errors caused by codewords from competitive clouds) are all pairwise independent. Thus, after taking the expectation w.r.t.\ the joint distribution of $(\mathbf{U}_{0},  \mathbf{X}_{00}, \mathbf{Y})$, we have
\begin{align}
 \bar{P}_{e}(R_{y},R_{z},n) 
&=  \mathbb{E} \Bigg[  \mathrm{Pr} \Bigg\{ \Bigg\{   \bigcup_{k=1}^{M_{y}-1} \mathcal{A}_{k}  \Bigg\} \bigcup \Bigg\{   \bigcup_{l=1}^{M_{z}-1} \mathcal{B}_{l}  \Bigg\}  \Bigg| \mathbf{U}_{0} = \bu_{0}, \mathbf{X}_{00} = \bx_{00}, \mathbf{Y} = \by     \Bigg\}  \Bigg]   \\
&\doteq  \mathbb{E} \Bigg[ \min  \Bigg\{ 1, \sum_{k=1}^{M_{y}-1} \mathrm{Pr}(\mathcal{A}_{k} | \mathbf{U}_{0} = \bu_{0}, \mathbf{X}_{00} = \bx_{00}, \mathbf{Y} = \by)  \nonumber  \\  & \;\;\;\;\;\;\;\;\;\;\;\;\;\;\;\;\;\;\;\;\;\;\;\;  +   \sum_{l=1}^{M_{z}-1} \mathrm{Pr}(\mathcal{B}_{l} | \mathbf{U}_{0} = \bu_{0}, \mathbf{X}_{00} = \bx_{00}, \mathbf{Y} = \by)  \Bigg\} \Bigg]    \\
&\doteq \mathbb{E} \bigg[ \min \bigg\{  1, e^{nR_{y}} \exp \Big\{-n \cdot   E_{0}(S, Q_{U_{0}Y} )    \Big\} \nonumber  \\  & \;\;\;\;\;\;\;\;\;\;\;\;\;\;\;\;\;\;\;\;\;\;\;\; +  e^{nR_{z}}   \exp \Big\{-n \cdot   E_{2}(S, Q_{Y} )    \Big\}   \bigg\} \bigg]    \\
&\doteq   \mathbb{E} \Bigg[ \min \Bigg\{  1,   \exp  \bigg\{ -n \cdot \min  \Big\{ \big[ E_{0}(S, Q_{U_{0}Y} ) -  R_{y} \big]   , 
\nonumber  \\  & \;\;\;\;\;\;\;\;\;\;\;\;\;\;\;\;\;\;\;\;\;\;\;\;~~~~~~~~~~~~~~~~~~~~~~~~  \big[ E_{2}(S, Q_{Y} ) -  R_{z} \big] \Big\}   \bigg\}   \Bigg\} \Bigg]   \\
&=   \mathbb{E} \Bigg[   \exp  \bigg( -n \cdot \min  \Big\{ \big[ E_{0}(S, Q_{U_{0}Y} ) -  R_{y} \big]_{+}   , \big[ E_{2}(S, Q_{Y} ) -  R_{z} \big]_{+} \Big\}   \bigg)  \Bigg]   \\
&\doteq   \exp \Bigg\{-n \cdot \min_{ Q_{Y|U_{0}X_{00}}} \Big[  D(Q_{Y|U_{0}X_{00}}||W_{Y|X_{00}}| P_{U_{0}X_{00}}  )  
\nonumber  \\  & \;\;\;\;\;\;\;\;\;\;\;\;\;\;\;\;\;\;\;\;\;\;\;\;~~~~~~~~~~~~~~~~~ + E_{3} \big( f(Q_{U_{0}X_{00}Y}), Q_{U_{0}Y}, R_{y}, R_{z} \big)  \Big]   \Bigg\}  ,
\end{align}
where we have defined
\begin{equation}
 E_{3}(S, Q_{U_{0}Y}, R_{y}, R_{z} )  \overset{\bigtriangleup}{=}   \min  \Big\{ \big[ E_{0}(S, Q_{U_{0}Y} ) -  R_{y} \big]_{+}   , \big[ E_{2}(S, Q_{Y} ) -  R_{z} \big]_{+} \Big\}.
\end{equation}

\subsection [A] {A Converse-Like\footnote{A converse result is usually w.r.t.\ both encoding and decoding. In our case, here and in Subsection 5.1, the converse results are w.r.t.\ the decoding only.} Result for the Strong User}
We have the following: \\ 
\textbf{Lemma 1.} For every empirical distribution $Q_{U_{0}X_{00}Y}$,
\begin{align}
\label{Lemma4Result}
  E_{3} \big( f(Q_{U_{0}X_{00}Y}), Q_{U_{0}Y} , R_{y}, R_{z} \big) \leq
\min  \bigg\{  \Big[  I_{Q}(U_{0};Y)  + & \Big[ I_{Q}(X_{00};Y|U_{0}) - R_{y} \Big]_{+} - R_{z}  \Big]_{+}, \nonumber \\ &\Big[ I_{Q}(X_{00};Y|U_{0}) - R_{y} \Big]_{+}    \bigg\}.
\end{align}
\\
\textbf{Proof.} We start by recalling that the function $E_{3}$ is defined as
\begin{align}
\label{minimumOFstrong}
 E_{3}(f(Q_{U_{0}X_{00}Y}), Q_{U_{0}Y}, R_{y}, R_{z} )  \overset{\bigtriangleup}{=}   \min  \Big\{ &\big[ E_{0}(f(Q_{U_{0}X_{00}Y}), Q_{U_{0}Y} ) -  R_{y} \big]_{+}   , \nonumber \\ 
&\big[ E_{2}( f(Q_{U_{0}X_{00}Y}) , Q_{Y} ) -  R_{z} \big]_{+} \Big\},
\end{align}
and we separately upper bound each one of the terms. We can upper bound them by choosing any specific distribution, instead of minimizing over them. Let us start with the left term:
\begin{align}
& \big[ E_{0}(f(Q_{U_{0}X_{00}Y}), Q_{U_{0}Y} ) -  R_{y} \big]_{+} \\
&= \min_{  \{ Q_{X|U_{0}Y} \in \mathcal{S}(Q_{U_{0}Y}): ~    f(Q_{U_{0}XY}) \geq f(Q_{U_{0}X_{00}Y})    \} } \big[ I_{Q}(X;Y|U_{0}) - R_{y} \big]_{+}  \\
&\leq  \big[ I_{Q}(X_{00};Y|U_{0}) - R_{y} \big]_{+}.
\end{align}
For the right term inside the minimum of (\ref{minimumOFstrong}), we have the following 
\begin{align}
& \big[ E_{2}(f(Q_{U_{0}X_{00}Y}), Q_{Y} ) -  R_{z} \big]_{+} \\
&= \min_{ \{ Q_{U'|Y} \in \mathcal{S}(Q_{Y})   \} }  \Big[  I_ {Q}(U';Y)  +   E_{1}( f(Q_{U_{0}X_{00}Y}), Q_{U'Y} )  -  R_{z}   \Big]_{+}   \\
&\leq   \Big[  I_ {Q}(U_{0};Y)  +   E_{1}(f(Q_{U_{0}X_{00}Y}), Q_{U_{0}Y} )  -  R_{z} \Big]_{+} \\
&=   \Bigg[  I_ {Q}(U_{0};Y)  +  \min_{ \{ Q_{X|U_{0}Y} \in \mathcal{S}(Q_{U_{0}Y}) : ~   f(Q_{U_{0}XY}) \geq f(Q_{U_{0}X_{00}Y})    \} } \big[ I_{Q}(X;Y|U_{0}) - R_{y} \big]_{+}   -  R_{z}   \Bigg]_{+} \\
&\leq   \Big[  I_ {Q}(U_{0};Y)  +  \big[ I_{Q}(X_{00};Y|U_{0}) - R_{y} \big]_{+}  -  R_{z}   \Big]_{+} .
\end{align}
Combining both upper bounds, we see that (\ref{Lemma4Result}) holds, thus completing the proof.  $\Box$

\subsection {An Optimal Universal Decoder}
Let us now select
\begin{align}
\label{OptimalFstrong}
  f(Q_{UXY}) =
I_{Q}(U;Y)  + [ I_{Q}(X;Y|U) - R_{y}]_{+}. 
\end{align}
We show that (\ref{OptimalFstrong}) achieves the maximum value of $E_{3} \big( f(Q_{U_{0}X_{00}Y}), Q_{U_{0}Y} , R_{y}, R_{z} \big)$, as given by Lemma 1, and therefore, this decoder has the same error exponent as the one of the optimal (ML) decoder. As before, we start with the left term inside the minimum of (\ref{minimumOFstrong}), and get
\begin{align}
& \big[ E_{0}(f(Q_{U_{0}X_{00}Y}), Q_{U_{0}Y} ) -  R_{y} \big]_{+} \\
&= \min_{  \{ Q_{X|U_{0}Y} \in \mathcal{S}(Q_{U_{0}Y})  \} } \Big\{ \big[ I_{Q}(X;Y|U_{0}) - R_{y} \big]_{+}  : ~    f(Q_{U_{0}XY}) \geq f(Q_{U_{0}X_{00}Y})  \Big\} \\
&= \min_{  \{ Q_{X|U_{0}Y} \in \mathcal{S}(Q_{U_{0}Y})  \} } \Big\{ \big[ I_{Q}(X;Y|U_{0}) - R_{y} \big]_{+}  : \nonumber    \\ &I_{Q}(U_{0};Y)  + [ I_{Q}(X;Y|U_{0}) - R_{y}]_{+} \geq I_{Q}(U_{0};Y)  + [ I_{Q}(X_{00};Y|U_{0}) - R_{y}]_{+}  \Big\} \\
&= \min_{  \{ Q_{X|U_{0}Y} \in \mathcal{S}(Q_{U_{0}Y})  \} } \Big\{ \big[ I_{Q}(X;Y|U_{0}) - R_{y} \big]_{+}  :  \nonumber    \\ &  [ I_{Q}(X;Y|U_{0}) - R_{y}]_{+} \geq  [ I_{Q}(X_{00};Y|U_{0}) - R_{y}]_{+}  \Big\} \\
\label{AAA}
&= [ I_{Q}(X_{00};Y|U_{0}) - R_{y}]_{+}.
\end{align}
For the right term inside the minimum of (\ref{minimumOFstrong}), 
\begin{align}
& \big[ E_{2}(f(Q_{U_{0}X_{00}Y}), Q_{Y} ) -  R_{z} \big]_{+} \\
&= \min_{ \{ Q_{UX|Y} \in \mathcal{S}(Q_{Y}) : ~   f(Q_{UXY}) \geq f(Q_{U_{0}X_{00}Y})  \} }   \Big[  I_ {Q}(U;Y)  +   \big[ I_{Q}(X;Y|U) - R_{y} \big]_{+}  -  R_{z}   \Big]_{+}  \\
&= \min_{ \{ Q_{UX|Y} \in \mathcal{S}(Q_{Y})   \} } \bigg\{  \Big[  I_ {Q}(U;Y)  +   \big[ I_{Q}(X;Y|U) - R_{y} \big]_{+}  -  R_{z}   \Big]_{+} : \nonumber    \\ &I_{Q}(U;Y)  + [ I_{Q}(X;Y|U) - R_{y}]_{+} \geq I_{Q}(U_{0};Y)  + [ I_{Q}(X_{00};Y|U_{0}) - R_{y}]_{+} \bigg\} \\
\label{BBB}
&= \Big[  I_ {Q}(U_{0};Y)  +   \big[ I_{Q}(X_{00};Y|U_{0}) - R_{y} \big]_{+}  -  R_{z}   \Big]_{+}.
\end{align}
Finally, compare the minimum between (\ref{AAA}) and (\ref{BBB}) to the right hand side of (\ref{Lemma4Result}).

\section {Universal Bin Index Decoding for the Weak User}

\subsection {Analysis for a General Decoding Metric and a Converse-Like Result}
Let us first derive the exact random coding error exponent of the following bin index decoder,
\begin{equation}
\label{GeneralDecoder}
  \hat{i}(\bz) = \operatorname*{arg\,max}_{0 \leq i \leq M_{z}-1} F(\bz,\bu_{i},\mathcal{C}_{i})  ,
\end{equation}
where
\begin{equation}
\label{GeneralDecoder2}
   F(\bz,\bu_{i},\mathcal{C}_{i})  \overset{\bigtriangleup}{=}  \frac {1}{M_{y}} \sum_{j=0}^{M_{y}-1} e^{nf(Q_{U_{i}X_{ij}Z})} ,
\end{equation}
and assume that $f$ is upper bounded by a real number $\Delta$. Note that (\ref{GeneralDecoder}) includes the optimal ML decoder (\ref{MLweak}) as a special case. 

To present the formula of $E^{*}_{z}(R_{y}, R_{z})$, the error exponent of (\ref{GeneralDecoder}), we first need a few definitions. For a given generic joint distribution $Q_{UXZ}$, let $I_{Q}(X;Z|U)$ denote the conditional mutual information between $X$ and $Z$ given $U$. For a given marginal $Q_{UZ}$, let $\mathcal{S}(Q_{UZ})$ denote the set of conditional distributions $\{Q_{X|UZ}\}$ such that $\sum_{z} Q_{UZ}(u,z)Q_{X|UZ}(x|u,z) = P_{UX}(u,x)$ for every $(u,x) \in \mathcal{U} \times \mathcal{X}$, where $P_{UX} = P_{U} \times P_{X|U}$. We first define
\begin{equation}
\label{Stage1}
  E_{1}(s, Q_{UZ})  =  \min_{Q_{X|UZ} \in \mathcal{S}(Q_{UZ})} \Big\{ [ I_{Q} (X;Z|U) - R_{y}]_{+}:f(Q_{UXZ}) +[R_{y}- I_{Q} (X;Z|U)]_{+} \geq s  \Big\},
\end{equation}
where $s$ is an arbitrary real. Next, for a given marginal $Q_{Z}$, define
\begin{equation}
\label{Stage2}
  E_{2}(s, Q_{Z})  =  \min_{Q_{U|Z} \in \mathcal{S}(Q_{Z})} \big[  I_{Q} (U;Z) +  E_{1}(s, Q_{UZ})  \big],
\end{equation}
where the minimization is across all $\{Q_{U|Z}\}$ such that $\sum_{z}Q_{Z}(z)Q_{U|Z}(u|z)=P_{U}(u)$ for every $u \in \mathcal{U}$. Finally, for a given $Q_{U_{0}Z}$, let
\begin{equation}
\label{S0Definition}
  s_{0}(Q_{U_{0}Z}) = R_{y} + \max_{\{ Q_{X|U_{0}Z} \in \mathcal{S}(Q_{U_{0}Z}): I_{Q} (X;Z|U_{0}) \leq R_{y} \}} \big[ f(Q_{U_{0}XZ}) - I_{Q} (X;Z|U_{0}) \big],
\end{equation}
and
\begin{equation}
  s_{1}(Q_{U_{0}X_{00}Z}) = \max \big\{ s_{0}(Q_{U_{0}Z}), f(Q_{U_{0}X_{00}Z})  \big\}.
\end{equation}
Now, the error exponent of the decoder (\ref{GeneralDecoder}) is given in the following lemma. \\
\textbf{Lemma 2.} Under the assumptions of Section 2, 
\begin{equation}
\label{Lemma1Result}
    E^{*}_{z}(R_{y}, R_{z}) = \min_{Q_{Z|U_{0}X_{00}}} \bigg\{ D(Q_{Z|U_{0}X_{00}}||W_{Z|X_{00}}| P_{U_{0}X_{00}}  ) +   \Big[E_{2}\big( s_{1}(Q_{U_{0}X_{00}Z}), Q_{Z} \big) - R_{z} \Big]_{+}   \bigg\},
\end{equation}
where $(U_{0},X_{00})$ is a replica of $(U,X)$, i.e., $P_{U_{0}X_{00}} = P_{UX}$. \\
\textbf{Proof.} The average probability of error, associated with (\ref{GeneralDecoder}), is given by
\begin{align}
\label{Union1}
  P_{e}^{*}  &=  \mathbb{E} \Bigg[ \mathrm{Pr} \Bigg\{ \bigcup_{i=1}^{M_{z}-1}  \Big\{ F(\mathbf{Z},\mathbf{U}_{i},\mathcal{C}_{i}) \geq  F(\mathbf{Z},\mathbf{U}_{0},\mathcal{C}_{0})  \Big\}   \Bigg\} \Bigg] \\
\label{Minimum1}  
 &\doteq   \mathbb{E} \Bigg[ \min \bigg \{1, M_{z} \cdot \mathrm{Pr} \Big\{ F(\mathbf{Z},\mathbf{U}_{1},\mathcal{C}_{1}) \geq  F(\mathbf{Z},\mathbf{U}_{0},\mathcal{C}_{0})  \Big\}   \bigg\} \Bigg]  ,  
\end{align}
where the expectation is w.r.t.\ the randomness of $\mathbf{U}_{0}$, $\mathcal{C}_{0}$ and $\mathbf{Z}$, where $\mathbf{Z}$ is the channel output in response to the input $\mathbf{X}_{00}$ (the transmitted codeword without loss of generality). The passage from (\ref{Union1}) to (\ref{Minimum1}) is due to the exponential tightness of the truncated union bound. 
Here, for a given $\bz$, $ \mathrm{Pr} \Big\{ F(\bz,\mathbf{U}_{1},\mathcal{C}_{1}) \geq  F(\bz,\bu_{0},\mathcal{C}_{0})  \Big\} $ is calculated w.r.t.\ the randomness of $\mathbf{U}_{1}$ and $\mathcal{C}_{1} = \big\{ \mathbf{X}_{1,0},...,\mathbf{X}_{1,(M_{y}-1)}  \big\}$, but for a given $\bu_{0}$ and $\mathcal{C}_{0}$. \\
Let $N_{1}(Q_{U_{1}X'Z})$ denote the number of codewords $\bx_{1,j} \in \mathcal{C}_{1}$, such that the joint empirical distribution of  $\bx_{1,j}$ with $(\bu_{1}, \bz)$ is $Q_{U_{1}X'Z}$, that is
\begin{align}
N_{1}(Q_{U_{1}X'Z}) = \sum_{j=0}^{M_{y}-1}  \mathcal{I} \Big\{ (\bu_{1}, \bx_{1,j}, \bz) \in \mathcal{T}(Q_{U_{1}X'Z}) \Big\}.
\end{align}
Defining
\begin{equation}
  s  \overset{\bigtriangleup}{=}  \frac {1}{n} \ln \Bigg [  \sum_{j=0}^{M_{y}-1}  e^{nf(Q_{U_{0}X_{0j}Z})}  \Bigg]  ,
\end{equation}
we have,
\begin{align}
   \mathrm{Pr} \Big\{ F(\bz,\bu_{1},\mathcal{C}_{1}) \geq  F(\bz,\bu_{0},\mathcal{C}_{0})  \Big\}  &=   \mathrm{Pr} \Big\{ M_{y} \cdot F(\bz,\bu_{1},\mathcal{C}_{1})   \geq  e^{ns}   \Big\}    \\
     &=   \mathrm{Pr} \Bigg\{  \sum_{j=0}^{M_{y}-1}  e^{nf(Q_{U_{1}X_{1j}Z})}   \geq  e^{ns}   \Bigg\}  \\
     &=   \mathrm{Pr} \Bigg\{  \sum_{Q_{X'|U_{1}Z} \in \mathcal{S}(Q_{U_{1}Z})} N_{1}(Q_{U_{1}X'Z})e^{nf(Q_{U_{1}X'Z})}    \geq  e^{ns}   \Bigg\}  \\
                                                        &\doteq   \mathrm{Pr} \Bigg\{  \max_{Q_{X'|U_{1}Z} \in \mathcal{S}(Q_{U_{1}Z})}  N_{1}(Q_{U_{1}X'Z})e^{nf(Q_{U_{1}X'Z})}    \geq  e^{ns}   \Bigg\}  \\
                                                        &=     \mathrm{Pr} \left\{  \bigcup_{Q_{X'|U_{1}Z} \in \mathcal{S}(Q_{U_{1}Z})} \Big\{N_{1}(Q_{U_{1}X'Z})e^{nf(Q_{U_{1}X'Z})}   \geq  e^{ns} \Big\}  \right\}  \\ 
                                                        &\doteq       \sum_{Q_{X'|U_{1}Z} \in \mathcal{S}(Q_{U_{1}Z})} \mathrm{Pr}  \Big\{ N_{1}(Q_{U_{1}X'Z})e^{nf(Q_{U_{1}X'Z})}  \geq  e^{ns} \Big\}   \\  
                                                       &\doteq       \max_{Q_{X'|U_{1}Z} \in \mathcal{S}(Q_{U_{1}Z})} \mathrm{Pr}  \Big\{ N_{1}(Q_{U_{1}X'Z})e^{nf(Q_{U_{1}X'Z})}  \geq  e^{ns} \Big\}.
\end{align}
Now, for a given $Q_{U_{1}X'Z}$, designating the joint empirical distribution of a randomly chosen $\bx'$ (given $\bu_{1}$) together with $(\bu_{1}, \bz)$, the binomial random variable $N_{1}(Q_{U_{1}X'Z})$ has $e^{nR_{y}}$ trials and probability of success which is of the exponential order of $e^{-n I_{Q}(X;Z|U)}$. Thus, a standard large deviations analysis (see, e.g., \cite[pp. 167--169]{MERHAV09}) yields
\begin{equation}
 \mathrm{Pr}  \bigg\{ N_{1}(Q_{U_{1}X'Z})  \geq  e^{n \big[ s-f(Q_{U_{1}X'Z}) \big] } \bigg\}  \doteq     e^{-nE_{0}(Q_{U_{1}X'Z})} ,
\end{equation}
where
\begin{align}
  E_{0}(Q_{U_{1}X'Z})  
                             &=             \left\{ 
               \begin{array}{l l}
   \big[ I_{Q}(X;Z|U) - R_{y} \big]_{+}   & \quad \text{  $f(Q_{U_{1}X'Z}) \geq s - \big[R_{y}- I_{Q}(X;Z|U) \big]_{+}$  }\\
                   \infty         & \quad \text{ $f(Q_{U_{1}X'Z})  <    s- \big[R_{y}- I_{Q}(X;Z|U) \big]_{+}.$  } 
               \end{array} \right.  
\end{align} 
Therefore, $ \max_{Q_{X'|U_{1}Z} \in \mathcal{S}(Q_{U_{1}Z})}  \mathrm{Pr}  \bigg\{ N_{1}(Q_{U_{1}X'Z})  \geq  e^{n \big[s-f(Q_{U_{1}X'Z}) \big] } \bigg\} $ decays according to 
\begin{equation}
  E_{1}(s, Q_{U_{1}Z} ) = \min_{Q_{X'|U_{1}Z} \in \mathcal{S}(Q_{U_{1}Z})}   E_{0}(Q_{U_{1}X'Z})  ,   
\end{equation}
which is given by (\ref{Stage1}). \\
The conditional pairwise error probability, given $\mathbf{U}_{1}=\bu_{1}$, is of the exponential order of $e^{-n E_{1}(s, Q_{U_{1}Z} ) }$. Averaging w.r.t.\ the randomness of $\mathbf{U}_{1}$, we get the exponential order of $e^{-n E_{2}(s, Q_{Z} ) }$, where $E_{2}(s, Q_{Z} )$ is defined as in (\ref{Stage2}). To see why this is true, consider the following:
\begin{align}
&  \sum_{\bu_{1} \in \mathcal{T} (P_{U})}  P_{U}(\bu_{1}) \cdot  \mathrm{Pr} \Big\{ F(\bz,\bu_{1},\mathcal{C}_{1}) \geq  F(\bz,\bu_{0},\mathcal{C}_{0})  \Big\} \\
  &\doteq \sum_{ Q_{U_{1}|Z} \in \mathcal{S}(Q_{Z}) }  \sum_{\bu_{1} \in \mathcal{T}(Q_{U_{1}|Z}|\bz)}  P_{U}(\bu_{1}) \cdot e^{-n \cdot E_{1}(s, Q_{U_{1}Z} )  }  \\
 &\doteq \sum_{Q_{U_{1}|Z} \in \mathcal{S}(Q_{Z}) }   e^{-n \cdot E_{1}(s, Q_{U_{1}Z} )  } \cdot e^{-n \cdot I_{Q}(U;Z)  }  \\
  &\doteq \max_{Q_{U_{1}|Z} \in \mathcal{S}(Q_{Z}) }   e^{-n \cdot \big[I_{Q}(U;Z) +  E_{1}(s, Q_{U_{1}Z} ) \big]  }  \\
				&=  e^{-n \cdot E_{2}(s, Q_{Z} )}.
\end{align}
Finally, we have that
\begin{align}
  P_{e}^{*}     &\doteq    \mathbb{E} \bigg[   \min \Big\{ 1,  M_{z}  \cdot   e^{-n \cdot E_{2}(S, Q_{Z} )  }  \Big\}  \bigg] \\
           &=  \mathbb{E} \bigg\{ e^{-n \big[ E_{2}(S, Q_{z} ) - R_{z} \big]_{+}   } \bigg\}  ,  
\end{align}
where the expectation is w.r.t.\ the randomness of
\begin{equation}
  S  =  \frac {1}{n} \ln \Bigg [  \sum_{j=0}^{M_{y}-1}  e^{nf(Q_{U_{0}X_{0j}Z})}  \Bigg]  ,
\end{equation}
the randomness of $Q_{Z}$, the empirical distribution of $\mathbf{Z}$, and $\mathbf{U}_{0}$, the real cloud center.\\
This expectation will be taken in two steps, the first is over the randomness of $\{ \mathbf{X}_{0,1},...,\mathbf{X}_{0,(M_{y}-1)}  \}$, while $\mathbf{X}_{00}=\bx_{00}$, $\mathbf{U}_{0}=\bu_{0}$ and $\mathbf{Z}=\bz$ are held fixed, whereas in the second step, the expectation is over the randomness of $\mathbf{X}_{00}$, $\mathbf{U}_{0}$ and $\mathbf{Z}$. 
Let $\bx_{00}$, $\bu_{0}$ and $\bz$ be given and let $\epsilon > 0$ be arbitrarily small. Then,
\begin{align}
 P_{e}^{*}( \bx_{00}, \bu_{0}, \bz  )  &\overset{\bigtriangleup}{=}  \mathbb{E} \bigg\{  e^{-n \big[ E_{2}(S, Q_{Z} ) - R_{z} \big]_{+} } \bigg|  \mathbf{X}_{00}=\bx_{00}, \mathbf{U}_{0}=\bu_{0}, \mathbf{Z}=\bz           \bigg\}  \\
  &\leq    \sum_{i}    \mathrm{Pr}  \Big\{  i\epsilon \leq S < (i+1)\epsilon \Big|  \mathbf{X}_{00}=\bx_{00}, \mathbf{U}_{0}=\bu_{0}, \mathbf{Z}=\bz  \Big\}  \nonumber  \\
 &~~~~~~~~~~~~~~~~~~~~~  \times  \exp \Big\{ -n \big[ E_{2}(i\epsilon, Q_{Z} ) - R_{z} \big]_{+}  \Big\}   , 
\end{align}
where $i$ ranges from $\frac{1}{\epsilon} f(Q_{U_{0}X_{00}Z } )$ to $(R_{y}+ \Delta )/\epsilon$.
Now,  
\begin{align}
        e^{nS}   &=    e^{n f(Q_{U_{0}X_{00}Z } ) }  +   \sum_{j=1}^{M_{y}-1}  e^{nf(Q_{U_{0}X_{0j}Z } ) } \\
 &=   e^{nf(Q_{U_{0}X_{00}Z } ) }   + \sum_{Q_{X'|U_{0}Z} \in \mathcal{S}(Q_{U_{0}Z})} N_{0}(Q_{U_{0}X'Z}) e^{nf(Q_{U_{0}X'Z})},
\end{align}
where $N_{0}(Q_{U_{0}X'Z})$ is the number of codewords in $\mathcal{C}_{0} \setminus \{\bx_{00}\}$, whose joint empirical distribution with $(\bu_{0},\bz)$ is $Q_{U_{0}X'Z}$.
On the one hand, we have
\begin{equation}
   \mathrm{Pr}  \Bigg\{ \sum_{Q_{X'|U_{0}Z} \in \mathcal{S}(Q_{U_{0}Z})} N_{0}(Q_{U_{0}X'Z}) e^{nf(Q_{U_{0}X'Z})} \geq e^{nt} \Bigg\}   
                                     \doteq         e^{-n \cdot E_{1}(t, Q_{U_{0}Z} )     }  ,
\end{equation}
and on the other hand, 
\begin{align}
   \mathrm{Pr}  \Bigg\{ \sum_{Q_{X'|U_{0}Z} \in \mathcal{S}(Q_{U_{0}Z})} &N_{0}(Q_{U_{0}X'Z}) e^{nf(Q_{U_{0}X'Z})} \leq e^{nt} \Bigg\} \nonumber    \\
&\doteq
     \mathrm{Pr}  \Bigg\{  \bigcap_{ Q_{X'|U_{0}Z} \in \mathcal{S}(Q_{U_{0}Z}) }   \Big\{  N_{0}(Q_{U_{0}X'Z})  \leq    e^{n[t- f(Q_{U_{0}X'Z}) ]}    \Big\}    \Bigg\}  .       
 \end{align}
This probability behaves exponentially like an indicator function of the condition that for every $Q_{X'|U_{0}Z} \in \mathcal{S}(Q_{U_{0}Z})$, either $I_{Q}(X;Z|U_{0}) \geq R_{y}$ or $R_{y} - I_{Q}(X;Z|U_{0}) \leq t-f(Q_{U_{0}X'Z})$ \cite{MERHAV2014}. I.e.,
\begin{align}
  \mathrm{Pr}  \Bigg\{ \sum_{Q_{X'|U_{0}Z} \in \mathcal{S}(Q_{U_{0}Z})} &N_{0}(Q_{U_{0}X'Z}) e^{nf(Q_{U_{0}X'Z})} \leq e^{nt} \Bigg\} \nonumber  \\  
&\doteq
  \mathcal{I}  \Bigg\{    R_{y}  \leq   \min_{ Q_{X'|U_{0}Z} \in \mathcal{S}(Q_{U_{0}Z}) }   \Big\{  I_{Q}(X;Z|U_{0})  +   [t - f(Q_{U_{0}X'Z}) ]_{+}      \Big\}    \Bigg\}  .       
 \end{align}
Let us now find what is the minimum value of $t$ for which the value of this indicator function is unity. The condition is equivalent to
\begin{equation}
                     \min_{ Q_{X'|U_{0}Z} \in \mathcal{S}(Q_{U_{0}Z}) }  \max_{0 \leq a \leq 1}   \bigg\{   I_{Q}(X;Z|U_{0})  + a  \Big[t- f(Q_{U_{0}X'Z}) \Big]     \bigg\}        \geq   R_{y} ,
 \end{equation}
or
\begin{equation}
                     \forall {Q_{X'|U_{0}Z} \in \mathcal{S}(Q_{U_{0}Z}) } \\\\\  \exists a \in [0,1]: \\\\\    I_{Q}(X;Z|U_{0})  + a  \Big[ t- f(Q_{U_{0}X'Z}) \Big]    \geq   R_{y} ,
 \end{equation}
or
\begin{equation}
                     \forall { Q_{X'|U_{0}Z} \in \mathcal{S}(Q_{U_{0}Z}) } \\\\\  \exists a \in [0,1]: \\\\\        t \geq   f(Q_{U_{0}X'Z}) + \frac{1}{a}  \Big( R_{y} - I_{Q}(X;Z|U_{0}) \Big)  ,
 \end{equation}
or, equivalently, 
\begin{align}
       t   &\geq   \max_{ Q_{X'|U_{0}Z} \in \mathcal{S}(Q_{U_{0}Z}) }  \min_{0 \leq a \leq 1}   \Bigg[    f(Q_{U_{0}X'Z})  + \frac{1}{a}  \Big( R_{y} -  I_{Q}(X;Z|U_{0}) \Big)   \Bigg]  \\
  &=   \max_{ Q_{X'|U_{0}Z} \in \mathcal{S}(Q_{U_{0}Z}) }    \Bigg[ f(Q_{U_{0}X'Z})  +       \left\{ 
               \begin{array}{l l}
                 R_{y} - I_{Q}(X;Z|U_{0})         & \quad \text{,  $R_{y} \geq I_{Q}(X;Z|U_{0})$  }\\
                  -  \infty                                     & \quad \text{,  $R_{y}    <  I_{Q}(X;Z|U_{0})$  } 
               \end{array} \right.    \Bigg]   \\
          &= R_{y} + \max_{\{ Q_{X'|U_{0}Z} \in \mathcal{S}(Q_{U_{0}Z}): \: I_{Q}(X;Z|U_{0}) \leq R_{y} \}} \Big[ f(Q_{U_{0}X'Z}) - I_{Q}(X;Z|U_{0}) \Big]  \\
          &\overset{\bigtriangleup}{=}    s_{0}(  Q_{U_{0}Z}   ).
\end{align}
Thus, in summary, we have
 \begin{align}
      \mathrm{Pr}  \Bigg\{   e^{nt}  \leq   \sum_{Q_{X'|U_{0}Z} \in \mathcal{S}(Q_{U_{0}Z})} &N_{0}(Q_{U_{0}X'Z}) e^{nf(Q_{U_{0}X'Z})}   \leq   e^{n(t+\epsilon)} \Bigg\}  \nonumber \\
 &\doteq 
                        \left\{ 
               \begin{array}{l l}
                0                             & \quad \text{$t <s_{0}(  Q_{U_{0}Z}   ) - \epsilon$  }\\
    e^{-n \cdot E_{1}(t, Q_{U_{0}Z} )     }        & \quad \text{$t \geq s_{0}( Q_{U_{0}Z} ) $  }       
               \end{array} \right.
\end{align}
Therefore, we get the expected error probability
\begin{align}
P_{e}^{*}(\bx_{00}, \bu_{0}, \bz )   
  &\leq   \sum_{i}    \mathrm{Pr}  \Bigg\{  e^{ni\epsilon}  \leq  \sum_{Q_{X'|U_{0}Z} \in \mathcal{S}(Q_{U_{0}Z})} N_{0}(Q_{U_{0}X'Z}) e^{nf(Q_{U_{0}X'Z})}   \leq      e^{n(i+1)\epsilon}  \Bigg\}       \nonumber \\
 &~~~~~~~~~~~~~~~~~~~
 \times \exp \Bigg\{-n \bigg[ E_{2} \Big(\max \big\{ i\epsilon, f(Q_{U_{0}X_{00}Z})  \big\}, Q_{Z} \Big) - R_{z} \bigg]_{+}   \Bigg\}     \\
      &\doteq     \sum_{i \geq s_{0}( Q_{U_{0}Z}) / \epsilon }  \exp \Big\{-n E_{1}(i\epsilon, Q_{U_{0}Z} )  \Big\}    \nonumber \\
 &~~~~~~~~~~~~~~~~~~~
 \times \exp \Bigg\{-n \bigg[ E_{2} \Big(\max \big\{ i\epsilon, f(Q_{U_{0}X_{00}Z})  \big\}, Q_{Z} \Big) - R_{z} \bigg]_{+}   \Bigg\} .
\end{align}
Since the dominant contribution to the sum over $i$ is due to the term $i = s_{0}(Q_{U_{0}Z}) / \epsilon$ (by the non-decreasing monotonicity of the functions $E_{1}(\cdot, Q_{U_{0}Z} )$ and $E_{2}(\cdot, Q_{Z} )$), we obtain
 \begin{align}
  P_{e}^{*}( \bx_{00}, \bu_{0}, \bz )  &\doteq  \exp \Bigg\{-n \bigg[ E_{2} \Big(\max \big\{ s_{0}(Q_{U_{0}Z} ), f(Q_{U_{0}X_{00}Z})  \big\}, Q_{Z} \Big) - R_{z}   \bigg]_{+}   \Bigg\}  \\
       &\overset{\bigtriangleup}{=}   \exp \Bigg\{-n \bigg[ E_{2} \Big( s_{1}(Q_{U_{0}X_{00}Z}) , Q_{Z} \Big) - R_{z} \bigg]_{+}   \Bigg\}.
\end{align}
Now, after taking the expectation w.r.t.\ the joint distribution of $(\mathbf{U}_{0},  \mathbf{X}_{00}, \mathbf{Z})$, we get the exact random coding error exponent (\ref{Lemma1Result}),
and the proof of Lemma 2 is complete.  $\Box$  \\
\\
Next, we introduce the following converse-like result for the weak user. \\
\textbf{Lemma 3.} For every empirical distribution $Q_{U_{0}X_{00}Z}$,
\begin{align}
\label{Lemma2eq}
  E_{2}(s_{1}(Q_{U_{0}X_{00}Z}), Q_{Z})  \leq
I_{Q}(U_{0};Z)  +  \big[ I_{Q}(X_{00};Z|U_{0}) - R_{y} \big]_{+}.
\end{align}
\textbf{Proof.} By (\ref{Stage2}),
\begin{align}
  E_{2}(s_{1}(Q_{U_{0}X_{00}Z}), Q_{Z})  &=   \min_{Q_{U|Z} \in \mathcal{S}(Q_{Z}) } \Big\{  I_{Q}(U;Z) +  E_{1}(s_{1}(Q_{U_{0}X_{00}Z}), Q_{UZ} )  \Big\} \\
&\leq   I_{Q}(U_{0};Z) +  E_{1}(s_{1}(Q_{U_{0}X_{00}Z}), Q_{U_{0}Z} )   ,
\end{align}
where
\begin{align}
 & E_{1}(s_{1}(Q_{U_{0}X_{00}Z}), Q_{U_{0}Z}) \nonumber  \\ 
&= \min_{ Q_{X'|U_{0}Z} \in \mathcal{S}(Q_{U_{0}Z}): ~ f(Q_{U_{0}X'Z}) +\big[R_{y}- I_{Q}(X';Z|U_{0}) \big]_{+} \geq s_{1}(Q_{U_{0}X_{00}Z})  } \big[ I_{Q}(X';Z|U_{0}) - R_{y} \big]_{+} .
\end{align} 
Now, since $s_{1}(Q_{U_{0}X_{00}Z})$ is given by the maximum
 \begin{align}
s_{1}(Q_{U_{0}X_{00}Z}) =    
 \max \Big\{ s_{0}(Q_{U_{0}Z}), f(Q_{U_{0}X_{00}Z})  \Big\} ,
\end{align}
we treat each case separately. First, if $s_{1}(Q_{U_{0}X_{00}Z}) =  f(Q_{U_{0}X_{00}Z})  $, 
\begin{align}
 & E_{1}( f(Q_{U_{0}X_{00}Z}), Q_{U_{0}Z}) \nonumber  \\ 
&= \min_{ Q_{X'|U_{0}Z} \in \mathcal{S}(Q_{U_{0}Z}): ~ f(Q_{U_{0}X'Z}) +\big[R_{y}- I_{Q}(X';Z|U_{0}) \big]_{+} \geq f(Q_{U_{0}X_{00}Z})  } \big[ I_{Q}(X';Z|U_{0}) - R_{y} \big]_{+}  \\
&\leq  \big[ I_{Q}(X_{00};Z|U_{0}) - R_{y} \big]_{+} ,
\end{align}
since the constraint is satisfied for $Q_{X_{00}|U_{0}Z} \in \mathcal{S}(Q_{U_{0}Z})$. 
\\
On the other hand, if $s_{1}(Q_{U_{0}X_{00}Z}) =  s_{0}(Q_{U_{0}Z}) $, which is given by (\ref{S0Definition}), we have
\begin{align}
 s_{0}(Q_{U_{0}Z}) &=   R_{y} +  f(Q_{U_{0}\tilde{X}Z}) - I_{Q}(\tilde{X};Z|U_{0}) ,
\end{align}
where we have denoted the maximizer of (\ref{S0Definition}) by $Q_{\tilde{X}|U_{0}Z}$, for which $I_{Q}(\tilde{X};Z|U_{0}) \leq R_{y}$ must be satisfied. 
Next, we upper bound the minimum defining $ E_{1}(s_{0}(Q_{U_{0}Z}), Q_{U_{0}Z})$ by using the same empirical distribution which is the maximizer of the right hand side of the constraint, for which the constraint becomes an exact equality:
\begin{align}
 & E_{1}(s_{0}(Q_{U_{0}Z}), Q_{U_{0}Z}) \\ 
&= \min_{ Q_{X'|U_{0}Z} \in \mathcal{S}(Q_{U_{0}Z}): ~ f(Q_{U_{0}X'Z}) +\big[R_{y}- I_{Q}(X';Z|U_{0}) \big]_{+} \geq s_{0}(Q_{U_{0}Z})  } \big[ I_{Q}(X';Z|U_{0}) - R_{y} \big]_{+}   \\
&= \min_{ Q_{X'|U_{0}Z} \in \mathcal{S}(Q_{U_{0}Z}): ~ f(Q_{U_{0}X'Z}) +\big[R_{y}- I_{Q}(X';Z|U_{0}) \big]_{+} \geq f(Q_{U_{0}\tilde{X}Z}) + R_{y}  - I_{Q}(\tilde{X};Z|U_{0})  } 
\big[ I_{Q}(X';Z|U_{0}) - R_{y} \big]_{+}   \\
&\leq \big[ I_{Q}(\tilde{X};Z|U) - R_{y} \big]_{+} =  0,
\end{align}
where the last equality is due to the constraint $I_{Q}(\tilde{X};Z|U) \leq R_{y}$. \\
Combining the last two upper bounds, we get 
\begin{align}
E_{1}(s_{1}(Q_{U_{0}X_{00}Z}), Q_{U_{0}Z}) 
&\leq  \big[ I_{Q}(X_{00};Z|U_{0}) - R_{y} \big]_{+},
\end{align}
and therefore
\begin{align}
  E_{2}(s_{1}(Q_{U_{0}X_{00}Z}), Q_{Z})  
\leq  I_{Q}(U_{0};Z)  +  \big[ I_{Q}(X_{00};Z|U_{0}) - R_{y} \big]_{+},
\end{align}
completing the proof of Lemma 3. $\Box$

\subsection {Analysis for a General Suboptimal Decoding Metric and a Direct Part}
Let us now derive the exact random coding error exponent of the following suboptimal bin index decoder:
\begin{equation}
\label{OrdinaryMLforWEAK}
  \tilde{i}(\bz) = \operatorname*{arg\,max}_{0 \leq i \leq M_{z}-1}
\Bigg\{ \max_{0 \leq j \leq M_{y}-1} f(Q_{U_{i}X_{ij}Z}) \Bigg\}  ,
\end{equation}
and assume, as before, that the function $f$ is upper bounded by $\Delta$.  \\
To present the formula of $\tilde{E}_{z}(R_{y}, R_{z})$, the error exponent of (\ref{OrdinaryMLforWEAK}), we first need a few new definitions. We first define
\begin{equation}
  \tilde{E}_{1}(t, Q_{UZ})  =  \min_{Q_{X|UZ} \in \mathcal{S}(Q_{UZ})} \Big\{ [ I_{Q} (X;Z|U) - R_{y}]_{+}:~f(Q_{UXZ}) \geq t  \Big\},
\end{equation}
where $t$ is an arbitrary real number. Next, for a given $Q_{Z}$, define
\begin{equation}
 \tilde{E}_{2}(t, Q_{Z})  =  \min_{Q_{U|Z} \in \mathcal{S}(Q_{Z})} [  I_{Q} (U;Z) +  \tilde{E}_{1}(t, Q_{UZ})  ].
\end{equation}
Finally, let
\begin{equation}
t_{0}(Q_{U_{0}Z}) = \max_{\{ Q_{X|U_{0}Z} \in \mathcal{S}(Q_{U_{0}Z}): I_{Q} (X;Z|U_{0}) \leq R_{y} \}}   f(Q_{U_{0}XZ}) ,
\end{equation}
and
\begin{equation}
  t_{1}(Q_{U_{0}X_{00}Z}) = \max \{ t_{0}(Q_{U_{0}Z}), f(Q_{U_{0}X_{00}Z})  \}.
\end{equation}
The error exponent of (\ref{OrdinaryMLforWEAK}) is given in the following lemma. \\
\textbf{Lemma 4.} Under the assumptions of Section 2, 
\begin{equation}
    \tilde{E}_{z}(R_{y}, R_{z}) = \min_{Q_{Z|U_{0}X_{00}}} \Big\{ D(Q_{Z|U_{0}X_{00}}||W_{Z|X_{00}}| P_{U_{0}X_{00}}  ) +   \big[ \tilde{E}_{2}(t_{1}(Q_{U_{0}X_{00}Z}), Q_{Z}) - R_{z} \big]_{+}   \Big\},
\end{equation}
where $(U_{0},X_{00})$ is a replica of $(U,X)$, i.e., $P_{U_{0}X_{00}} = P_{UX}$. \\
\\
\textbf{Proof.} The average probability of error, associated with (\ref{OrdinaryMLforWEAK}), is given by
\begin{equation}
  P_{e}^{*}   \doteq   \mathbb{E} \left[ \min \left \{ 1, M_{z} \cdot
\mathbb{E}  \Bigg( \min \bigg[ 1, M_{y} \cdot 
\mathrm{Pr} \Big\{   f(Q_{U_{1}X_{10}Z})  \geq  \max_{\bx \in \mathcal{C}_{0}}  f(Q_{U_{0}XZ})   \Big\}   \bigg]  \Bigg)  \right\}  \right]   ,  
\end{equation}
where the inner expectation is w.r.t.\ the randomness of $\mathbf{U}_{1}$, the outer expectation is w.r.t. the randomness of $\mathbf{U}_{0}$, $\mathcal{C}_{0}$ and $\mathbf{Z}$, the latter being the channel output in response to $\mathbf{X}_{00}$. \\
To see why this is true, observe that the average error probability, $\bar{P}_{e}(R_{y},R_{z},n)$, associated with (\ref{OrdinaryMLforWEAK}), is defined as 
\begin{align}
 &\bar{P}_{e}(R_{y},R_{z},n)  \nonumber \\    
&\overset{\bigtriangleup}{=}  \frac{1}{M_{y} M_{z}} \sum_{i=0}^{M_{z}-1} \sum_{j=0}^{M_{y}-1} \mathrm{Pr} \Bigg\{   \bigcup_{l \neq i} \bigcup_{k} \Big\{ f(Q_{U_{l}X_{lk}Z})  \geq  \max_{\bx \in \mathcal{C}_{i}}  f(Q_{U_{i}XZ})  \Big| \mathbf{X}_{ij} \; \mathrm{sent}   \Big\}        \Bigg\} 
\end{align}
where $\mathrm{Pr}\{\cdot\}$ pertains to the randomness of the codebook as well as that of the channel output given its input. 
We define the following unions of events                                      
\begin{equation}
   \mathcal{G} \overset{\bigtriangleup}{=} \bigcup_{l=1}^{M_{z}-1} \mathcal{G}_{l}  \overset{\bigtriangleup}{=} \bigcup_{l=1}^{M_{z}-1} \bigcup_{k=0}^{M_{y}-1} \mathcal{G}_{lk}    \overset{\bigtriangleup}{=} \bigcup_{l=1}^{M_{z}-1} \bigcup_{k=0}^{M_{y}-1} \Big\{ f(Q_{U_{l}X_{lk}Z})  \geq  \max_{\bx \in \mathcal{C}_{0}} f(Q_{U_{0}XZ})  \Big\} .
\end{equation}
\\
Define 
\begin{equation}
t  \overset{\bigtriangleup}{=}   \max_{\bx \in \mathcal{C}_{0}}   f(Q_{U_{0}XZ})  .
\end{equation}
The probability of $\mathcal{G}_{lk}$, conditioned on $\mathbf{U}_{l}$, is given by
\begin{align}
 \mathrm{Pr}(\mathcal{G}_{lk} | \mathbf{U}_{l} = \bu')   &\overset{\bigtriangleup}{=} 
	\mathrm{Pr} \bigg\{ f(Q_{U_{l}X_{lk}Z})  \geq  \max_{\bx \in \mathcal{C}_{0}} 
 f(Q_{U_{0}XZ})  \bigg| \mathbf{U}_{l} = \bu' \bigg\} \\
\label{GUMP1c}
&=     \sum_{\{\bx':~  f(Q_{U'X'Z}) \geq t   \}} P ( \bx' | \bu' ) \\
\label{GUMP1d}
&\doteq  \max_{\{ Q_{X'|U'Z} \in \mathcal{S}(Q_{U'Z}) : ~  f(Q_{U'X'Z})  \geq t  \}} \exp \Big\{-n \cdot  
I_ {Q}(X';Z|U')   \Big\}  \\
&=   \exp \Bigg\{-n \cdot  \min_{\{ Q_{X'|U'Z} \in \mathcal{S}(Q_{U'Z}) : ~  f(Q_{U'X'Z})  \geq t  \}}     I_ {Q}(X';Z|U')     \Bigg\}  \\
&\overset{\bigtriangleup}{=}   \exp \Big\{-n \cdot   \tilde{E}_{0}(t, Q_{U'Z} )    \Big\},
\end{align}
where the passage from (\ref{GUMP1c}) to (\ref{GUMP1d}) is due to (\ref{GUMP1a})-(\ref{GUMP1b}). For a given $\mathbf{U}_{l} = \bu'$, the events $\{\mathcal{G}_{lk}\}_{k}$ are all pairwise independent since we have assumed that the various codewords are pairwise conditional independent given the cloud center. Using the exponential tightness of the truncated union bound, we get
\begin{align}
 \mathrm{Pr} \Bigg\{  \bigcup_{k=0}^{M_{y}-1} \mathcal{G}_{lk} \Bigg| \mathbf{U}_{l} = \bu'   \Bigg\}   
&\doteq   \min  \Bigg\{ 1, \sum_{k=0}^{M_{y}-1} \mathrm{Pr}(\mathcal{G}_{lk} | \mathbf{U}_{l} = \bu')  \Bigg\}  \\  
&=   \min  \Big\{ 1, M_{y} \cdot  \mathrm{Pr}(\mathcal{G}_{l,0} | \mathbf{U}_{l} = \bu')  \Big\}  \\
&\doteq   \min  \bigg\{ 1, e^{nR_{y}} \cdot  \exp \Big\{-n \cdot   \tilde{E}_{0}(t, Q_{U'Z} )    \Big\}  \bigg\} \\
&\overset{\bigtriangleup}{=}   \exp \Big\{-n \cdot   \tilde{E}_{1}(t, Q_{U'Z}  )    \Big\}  ,
\end{align}
where
\begin{equation}
  \tilde{E}_{1}(t, Q_{U'Z} ) =  \min_{Q_{X'|U'Z} \in \mathcal{S}(Q_{U'Z})} \Big\{\big[ I_ {Q}(X';Z|U') - R_{y} \big]_{+}: \:    f(Q_{U'X'Z})  \geq t  \Big\}.
\end{equation}
\\
Next, we obtain the probability of $\mathcal{G}_{l}$ by calculating the expectation w.r.t.\ the randomness of $\mathbf{U}_{l}$:
\begin{align}
\mathrm{Pr} \big\{   \mathcal{G}_{l}   \big\}  &=  \sum_{\bu' \in \mathcal{T}(P_{U})}  P_{U}(\bu')  \cdot  
 \mathrm{Pr} \Bigg\{  \bigcup_{k=0}^{M_{y}-1} \mathcal{G}_{lk} \Bigg| \mathbf{U}_{l} = \bu'   \Bigg\}   \\
\label{GUMP2c}
&\doteq \sum_{\bu' \in \mathcal{T}(P_{U})}    P_{U}(\bu')  \cdot  
  \exp \Big\{-n \cdot   \tilde{E}_{1}(t, Q_{U'Z} )    \Big\}   \\
\label{GUMP2d}
&\doteq   \exp \Bigg\{-n \cdot \min_{ \{ Q_{U'|Z} \in \mathcal{S}(Q_{Z})   \} } 
\Big[ I_ {Q}(U;Z)  +   \tilde{E}_{1}(t, Q_{U'Z} )   \Big]  \Bigg\} \\
&\overset{\bigtriangleup}{=}   \exp \Big\{-n \cdot   \tilde{E}_{2}(t, Q_{Z} )    \Big\} ,
\end{align} 
where the passage from (\ref{GUMP2c}) to (\ref{GUMP2d}) is due to (\ref{GUMP2a})-(\ref{GUMP2b}).
Conditioning on $\mathbf{U}_{0}$, $\mathcal{C}_{0}$ and $\mathbf{Z}$, the events $\{\mathcal{G}_{l}\}$ are all pairwise independent since the various cloud centers are all independent. We get
\begin{align}
 \mathrm{Pr} \Bigg\{  \bigcup_{l=1}^{M_{z}-1} \mathcal{G}_{l} \Bigg| \mathbf{U}_{0}, \mathcal{C}_{0}, \mathbf{Z}   \Bigg\}   
&\doteq   \min  \Bigg\{ 1, \sum_{l=1}^{M_{z}-1} \mathrm{Pr}(\mathcal{G}_{l} | \mathbf{U}_{0}, \mathcal{C}_{0}, \mathbf{Z} ) \Bigg\}  \\  
&=   \min  \Big\{ 1, (M_{z}-1) \cdot  \mathrm{Pr}(\mathcal{G}_{1} | \mathbf{U}_{0}, \mathcal{C}_{0}, \mathbf{Z} ) \Big\}  \\
&\doteq   \min  \bigg\{ 1, e^{nR_{z}} \cdot  \exp \Big\{-n \cdot   \tilde{E}_{2}(t, Q_{Z} )    \Big\}  \bigg\} \\
&=   \exp \Big\{-n \cdot  \big[  \tilde{E}_{2}(t, Q_{Z} )  - R_{z}  \big]_{+}    \Big\} .
\end{align}
Finally, we have that
\begin{equation}
  \tilde{P}_{e}   =   \mathbb{E} \bigg[ \exp \Big\{-n \cdot  \big[  \tilde{E}_{2}(T, Q_{Z} )  - R_{z}  \big]_{+}    \Big\} \bigg]  ,  
\end{equation}
where the expectation is taken w.r.t.\ the randomness of 
\begin{equation}
T  =   \max_{\bx \in \mathcal{C}_{0}}   f(Q_{U_{0}XZ})  ,
\end{equation}
and the randomness of $Q_{Z}$ and $U_{0}$, the correct cloud center. 
This expectation will be taken in two steps, first, over the randomness of $\{ \mathbf{X}_{0,1},...,\mathbf{X}_{0,(M_{y}-1)}  \}$, while $\mathbf{X}_{00}=\bx_{00}$, $\mathbf{U}_{0}=\bu_{0}$ and $\mathbf{Z}=\bz$ are held fixed, and then - over the randomness of $\mathbf{X}_{00}$, $\mathbf{U}_{0}$ and $\mathbf{Z}$. Let $\bx_{00}$, $\bu_{0}$ and $\bz$ be given and let $\epsilon > 0$ be arbitrarily small. Then,
\begin{align}
 \tilde{P}_{e}( \bx_{00}, \bu_{0}, \bz  )  &\overset{\bigtriangleup}{=}  \mathbb{E} \bigg\{  e^{-n \big[ \tilde{E}_{2}(T, Q_{Z} ) - R_{z} \big]_{+} } \bigg|  \mathbf{X}_{00}=\bx_{00}, \mathbf{U}_{0}=\bu_{0}, \mathbf{Z}=\bz           \bigg\}  \\
   &\leq    \sum_{i}    \mathrm{Pr}  \Big\{  i\epsilon \leq  T  < (i+1)\epsilon \Big|  \mathbf{X}_{00}=\bx_{00}, \mathbf{U}_{0}=\bu_{0}, \mathbf{Z}=\bz  \Big\}   \nonumber  \\
 &~~~~~~~~~~~~~~~~~~~~~  \times   \exp \Big\{ -n \big[ \tilde{E}_{2}(i\epsilon, Q_{Z} ) - R_{z} \big]_{+}  \Big\}   , 
\end{align}
where $i$ ranges from $\frac{1}{\epsilon} f(Q_{U_{0}X_{00}Z } )$ to $\Delta /\epsilon$.
Now,  
\begin{align}
    T  =   \max \bigg\{  f(Q_{U_{0}X_{00}Z } )  ,  \max_{1 \leq j \leq M_{y}-1}  f(Q_{U_{0}X_{0j}Z } ) \bigg\}.
\end{align}
On the one hand, we have:
\begin{align}
   \mathrm{Pr}  \bigg\{  \max_{1 \leq j \leq M_{y}-1}  f(Q_{U_{0}X_{0j}Z } )  \geq   t   \bigg\}     
&= \mathrm{Pr} \Bigg\{   \bigcup_{1 \leq j \leq M_{y}-1}  \Big\{  f(Q_{U_{0}X_{0j}Z } )  \geq   t   \Big\} \Bigg\}  \\
 &\doteq  \min \Bigg\{1,  (M_{y} - 1)  \cdot  \mathrm{Pr}  \Big\{  f(Q_{U_{0}X_{01}Z } )  \geq  t \Big\}  \Bigg\}  \\
   &\doteq  \min \bigg\{1,  e^{nR_{y}}  \cdot    \exp \Big\{-n \cdot \tilde{E}_{0}(t, Q_{U_{0}Z}) \Big\} \bigg\}     \\
		   &=     \exp \Big\{ -n [ \tilde{E}_{0}(t, Q_{U_{0}Z}) - R_{y} ]_{+}   \Big\}   \\ 
		  &=     \exp \Big\{ -n \cdot \tilde{E}_{1}(t, Q_{U_{0}Z})     \Big\} .
\end{align}
On the other hand,
\begin{align}
   \mathrm{Pr} \bigg\{  \max_{1 \leq j \leq M_{y}-1}  f(Q_{U_{0}X_{0j}Z } )  <   t   \bigg\}    
  &=   \mathrm{Pr}  \Bigg\{  \bigcap_{1 \leq j \leq M_{y}-1 }   \Big\{  f(Q_{U_{0}X_{0j}Z } )  <   t   \Big\}    \Bigg\}              \\
   &=     \bigg [  \mathrm{Pr} \Big\{  f(Q_{U_{0}X_{01}Z } )  <   t   \Big\}   \bigg] ^{M_{y} - 1}    \\
    &\doteq      \bigg [ 1  -     e^{-n \cdot \tilde{E}_{0}(t, Q_{U_{0}Z})}    \bigg]  ^  {e^{nR_{y}} }    \\
   &=    \exp \Bigg\{      e^{nR_{y}}  \cdot    \ln    \bigg [ 1 -  e^{-n \cdot \tilde{E}_{0}(t, Q_{U_{0}Z})}    \bigg]  \Bigg\}    \\
   &\doteq    \exp \Big\{  -   e^{n \cdot [  R_{y} -  \tilde{E}_{0}(t, Q_{U_{0}Z}) ]   }   \Big\}    \\ 
 &\doteq   \left\{ 
               \begin{array}{l l}
                0,        & \quad \text{$R_{y} >  \tilde{E}_{0}(t, Q_{U_{0}Z})$  }\\
                1,        & \quad \text{$R_{y} <  \tilde{E}_{0}(t, Q_{U_{0}Z})$  },
               \end{array} \right.
\end{align}
which can also be written as:
 \begin{equation}
\mathrm{Pr} \bigg\{  \max_{1 \leq j \leq M_{y}-1}  f(Q_{U_{0}X_{0j}Z } )  <   t   \bigg\}  
           \doteq \mathcal{I} \Big\{  R_{y} <  \tilde{E}_{0}(t, Q_{U_{0}Z})  \Big\}    .
\end{equation}
Let us now find the minimum $t$ for which the value of this indicator function is unity. The condition is equivalent to
\begin{equation}
                     \min_{ Q_{X'|U_{0}Z} \in \mathcal{S}(Q_{U_{0}Z}) }  \max_{0 \leq a < \infty}   \bigg\{   I_{Q}(X';Z|U_{0})  + a  \Big[t- f(Q_{U_{0}X'Z}) \Big]     \bigg\}        \geq   R_{y} ,
 \end{equation}
or
\begin{equation}
                     \forall {Q_{X'|U_{0}Z} \in \mathcal{S}(Q_{U_{0}Z}) } ~  \exists a \in [0, \infty ): ~   I_{Q}(X';Z|U_{0})  + a  \Big[ t- f(Q_{U_{0}X'Z}) \Big]    \geq   R_{y} ,
 \end{equation}
or
\begin{equation}
                     \forall { Q_{X'|U_{0}Z} \in \mathcal{S}(Q_{U_{0}Z}) } ~  \exists a \in [0, \infty ): ~      t \geq   f(Q_{U_{0}X'Z}) + \frac{1}{a}  \Big( R_{y} - I_{Q}(X';Z|U_{0}) \Big)  ,
 \end{equation}
or, equivalently, 
\begin{align}
       t   &\geq   \max_{ Q_{X'|U_{0}Z} \in \mathcal{S}(Q_{U_{0}Z}) }  \min_{0 \leq a < \infty}   \Bigg[    f(Q_{U_{0}X'Z})  + \frac{1}{a}  \Big( R_{y} -  I_{Q}(X';Z|U_{0}) \Big)   \Bigg]  \\
  &=   \max_{ Q_{X'|U_{0}Z} \in \mathcal{S}(Q_{U_{0}Z}) }    \Bigg[ f(Q_{U_{0}X'Z})  +       \left\{ 
               \begin{array}{l l}
                   0     ,                      & \quad \text{$R_{y} \geq I_{Q}(X';Z|U_{0})$  }\\
                  -  \infty ,                  & \quad \text{$R_{y}    <  I_{Q}(X';Z|U_{0})$  } 
               \end{array} \right.    \Bigg]   \\
          &=  \max_{\{ Q_{X'|U_{0}Z} \in \mathcal{S}(Q_{U_{0}Z}): \: I_{Q}(X';Z|U_{0}) \leq R_{y} \}}    f(Q_{U_{0}X'Z})     \\
          &\overset{\bigtriangleup}{=}    t_{0}(  Q_{U_{0}Z}   ).
\end{align}
Thus, in summary, we have
 \begin{align}
      \mathrm{Pr}  \bigg\{   t  \leq   \max_{1 \leq j \leq M_{y}-1}  f(Q_{U_{0}X_{0j}Z } )  \leq   t+\epsilon \bigg\}  
 \doteq 
                        \left\{ 
               \begin{array}{l l}
                0                             & \quad \text{,  $t <t_{0}(  Q_{U_{0}Z}   ) - \epsilon$  }\\
    e^{-n \cdot \tilde{E}_{1}(t, Q_{U_{0}Z} )     }        & \quad \text{,  $t \geq t_{0}( Q_{U_{0}Z} ) $  }       
               \end{array} \right.
\end{align}
Then, the expected error probability w.r.t.\ $\{ \mathbf{X}_{0,1},...,\mathbf{X}_{0,(M_{y}-1)}  \}$ yields
\begin{align}
  \tilde{P}_{e}( \bx_{00}, \bu_{0}, \bz  )   
     &\doteq   \sum_{i}    \mathrm{Pr}  \bigg\{ i\epsilon  \leq   \max_{1 \leq j \leq M_{y}-1}  f(Q_{U_{0}X_{0j}Z } )    <         (i+1)\epsilon  \bigg\}      \nonumber \\
 &\qquad    \times  \exp \bigg\{ -n \Big[ \tilde{E}_{2}\big(\max \{f(Q_{U_{0}X_{00}Z}), i\epsilon \}, Q_{Z} \big) - R_{z} \Big]_{+}   \bigg\}     \\
      &\doteq     \sum_{i \geq t_{0}(Q_{U_{0}Z}) / \epsilon }  \exp \Big\{-n \tilde{E}_{1}(i\epsilon, Q_{U_{0}Z})  \Big\}    \nonumber \\
 &\qquad    \times  \exp \bigg\{ -n \Big[ \tilde{E}_{2}\big(\max \{f(Q_{U_{0}X_{00}Z}), i\epsilon \}, Q_{Z} \big) - R_{z} \Big]_{+}   \bigg\}  .
\end{align}
Since the dominant contribution to the sum over $i$ is due to the term $i = t_{0}(Q_{U_{0}Z}) / \epsilon$ (by the non-decreasing monotonicity of the functions $\tilde{E}_{1}(\cdot, Q_{U_{0}Z} )$ and $\tilde{E}_{2}(\cdot, Q_{Z} )$), we obtain
 \begin{align}
  \tilde{P}_{e}( \bx_{00}, \bu_{0}, \bz )  &\doteq  \exp \Bigg\{-n \bigg[ \tilde{E}_{2} \Big(\max \big\{ t_{0}(Q_{U_{0}Z} ), f(Q_{U_{0}X_{00}Z})  \big\}, Q_{Z} \Big) - R_{z}   \bigg]_{+}   \Bigg\}  \\
       &\overset{\bigtriangleup}{=}   \exp \Bigg\{-n \bigg[ \tilde{E}_{2} \Big( t_{1}(Q_{U_{0}X_{00}Z}) , Q_{Z} \Big) - R_{z} \bigg]_{+}   \Bigg\}.
\end{align}
Now, after taking the expectation w.r.t.\ the joint distribution of $(\mathbf{U}_{0},  \mathbf{X}_{00}, \mathbf{Z})$, we get
\begin{equation}
    \tilde{E}_{z}(R_{y}, R_{z}) = \min_{Q_{Z|U_{0}X_{00}}} \Big\{ D(Q_{Z|U_{0}X_{00}}||W_{Z|X_{00}}| P_{U_{0}X_{00}}  ) +   \big[ \tilde{E}_{2}(t_{1}(Q_{U_{0}X_{00}Z}), Q_{Z}) - R_{z} \big]_{+}   \Big\},
\end{equation}
and the proof of Lemma 4 is complete. $\Box$ \\
Let us now select
\begin{align}
\label{AgoodFunction}
  f(Q_{UXZ}) =
I_{Q}(U;Z)  + [ I_{Q}(X;Z|U) - R_{y}]_{+}. 
\end{align}
We show that (\ref{AgoodFunction}) achieves the maximum of $ E_{2}(s_{1}(Q_{U_{0}X_{00}Z}), Q_{Z})$, as given by Lemma 3, and therefore, this decoder has the same error exponent as that of the optimal decoder. First, the threshold $t_{0}(Q_{U_{0}Z}) $ can be easily simplified as
\begin{align}
 t_{0}(Q_{U_{0}Z})   &=  \max_{\{ Q_{X|U_{0}Z} \in \mathcal{S}(Q_{U_{0}Z}): I_{Q} (X;Z|U_{0}) \leq R_{y} \}}  f(Q_{U_{0}XZ})  \\
&=  \max_{\{ Q_{X|U_{0}Z} \in \mathcal{S}(Q_{U_{0}Z}): I_{Q} (X;Z|U_{0}) \leq R_{y} \}} 
\Big\{ I_{Q}(U_{0};Z)  + [ I_{Q}(X;Z|U_{0}) - R_{y}]_{+}  \Big\} \\
&=  I_{Q}(U_{0};Z)  +  \max_{\{ Q_{X|U_{0}Z} \in \mathcal{S}(Q_{U_{0}Z}): I_{Q} (X;Z|U_{0}) \leq R_{y} \}}  [ I_{Q}(X;Z|U_{0}) - R_{y}]_{+}  \\
&=  I_{Q}(U_{0};Z) .
\end{align}
Now, $t_{1}(Q_{U_{0}X_{00}Z})$ is given by 
\begin{align}
t_{1}(Q_{U_{0}X_{00}Z}) &= 
 \max \Big\{ I_{Q}(U_{0};Z),~  I_{Q}(U_{0};Z)  + [ I_{Q}(X_{00};Z|U_{0}) - R_{y}]_{+}   \Big\} \\
&= I_{Q}(U_{0};Z)  + [ I_{Q}(X_{00};Z|U_{0}) - R_{y}]_{+}.
\end{align}
In general, the constraint of the inner minimization problem defining $ \tilde{E}_{2}(t_{1}(Q_{U_{0}X_{00}Z}), Q_{Z})$ is given by
\begin{align}
f(Q_{UXZ})  \geq  t_{1}(Q_{U_{0}X_{00}Z}),
\end{align}
which can now be written as
\begin{align}
I_{Q}(U;Z)  + [ I_{Q}(X;Z|U) - R_{y}]_{+}  \geq I_{Q}(U_{0};Z)  + [ I_{Q}(X_{00};Z|U_{0}) - R_{y}]_{+},
\end{align} 
or simply by $f(Q_{UXZ})  \geq f(Q_{U_{0}X_{00}Z})$. Eventually, we have the following
\begin{align}
  \tilde{E}_{2}(t_{1}(Q_{U_{0}X_{00}Z}), Q_{Z}) 
&=   \min_{Q_{UX|Z} \in \mathcal{S}(Q_{Z}):~ f(Q_{UXZ})  \geq f(Q_{U_{0}X_{00}Z}) }   f(Q_{UXZ})     \\
&= I_{Q}(U_{0};Z)  + [ I_{Q}(X_{00};Z|U_{0}) - R_{y}]_{+},
\end{align}
which is the same expression as on the right hand side of (\ref{Lemma2eq}).

\section {Gallager-Style Lower Bounds}
In this section, we prove Theorem 5. 

\subsection {Derivation of eq.\ (\ref{StrongTerm1})}
We start by changing the clipping operator to a maximization problem and using convexity properties to change the order of the maximization and the minimization:
\begin{align}
\label{ChangeOrder1}
\hat{E}_{y,1}( R_{y} ) 
&\overset{\bigtriangleup}{=} \min_{ V } \bigg\{  D( V \| W |  P   )  + \Big[ I(X;Y|U) - R_{y} \Big]_{+}    \bigg\} \\
&=  \min_{ V } \bigg\{  D( V \| W |  P   )  +  \max_{\rho \in [0,1] } \Big\{ \rho \cdot  \big[ I(X;Y|U) - R_{y} \big] \Big\}    \bigg\} \\
\label{ChangeOrder2}
&=  \max_{\rho \in [0,1] } \bigg\{ - \rho R_{y}  +  \min_{ V }  \Big\{  D( V \| W |  P   )  +   \rho \cdot  I(X;Y|U)  \Big\}    \bigg\}.
\end{align}
Next,
\begin{align}
& \min_{ V }  \Big\{  D( V \| W |  P   )  +   \rho \cdot  I(X;Y|U)  \Big\}  \nonumber  \\
&= \min_{ V,Q }  \Bigg\{  \sum_{x,u}P(x,u)\sum_{y} V(y|x,u)\log \frac{V(y|x,u)}{W(y|x)}   \nonumber \\  & ~~~~~~~~~~~~~~~~~~~~~~~~~~~~~~~~+ \rho \sum_{x,u,y} P(x,u)V(y|x,u) \log \frac{V(y|x,u)}{Q(y|u)}  \Bigg\} \\
&= \min_{ V,Q }  \Bigg\{  \sum_{x,u,y} P(x,u)V(y|x,u) 
 \Bigg[  \log \frac{V(y|x,u)}{W(y|x)}+ \rho \log \frac{V(y|x,u)}{Q(y|u)} \Bigg]  \Bigg\} \\
\label{BeforeMinV}
&= \min_{ V,Q }  \Bigg\{  \sum_{x,u,y} P(x,u)V(y|x,u) 
   \log \Bigg[ \frac{V^{1+\rho}(y|x,u)}{W(y|x)Q^{\rho}(y|u)}  \Bigg]  \Bigg\}.
\end{align}
First, we minimize over the auxiliary channel $V$. Holding the auxiliary channel $Q$ fixed, and differentiating w.r.t.\ $V(y|x,u)$, we find that the minimizing distribution is given by
\begin{align}
V^{*}(y|x,u) = \frac{ W^{\frac{1}{1+\rho}}(y|x)Q^{\frac{\rho}{1+\rho}}(y|u) }{\sum_{y'} W^{\frac{1}{1+\rho}}(y'|x)Q^{\frac{\rho}{1+\rho}}(y'|u) }. 
\end{align}
Substituting it back into (\ref{BeforeMinV}) and summing over $y$, we get that
\begin{align}
 \min_{ Q }  \Bigg\{  \sum_{x,u,y} &P(x,u)V^{*}(y|x,u) 
   \log \Bigg[ \frac{ [V^{*}(y|x,u)]^{1+\rho} }{W(y|x)Q^{\rho}(y|u)}  \Bigg]  \Bigg\} \nonumber  \\ 
&=  \min_{ Q }  \Bigg\{ -(1+\rho) \sum_{x,u} P(x,u) 
   \log \Bigg[ \sum_{y} W^{\frac{1}{1+\rho}}(y|x)Q^{\frac{\rho}{1+\rho}}(y|u)  \Bigg]  \Bigg\} \\
&=  \min_{ Q }  \Bigg\{ -(1+\rho) \sum_{u} P(u)  \sum_{x} P(x|u) 
   \log \Bigg[ \sum_{y} W^{\frac{1}{1+\rho}}(y|x)Q^{\frac{\rho}{1+\rho}}(y|u)  \Bigg]  \Bigg\} \\
\label{JENSEN1}
&\geq  \min_{ Q }  \Bigg\{ -(1+\rho) \sum_{u} P(u)   
   \log \Bigg[ \sum_{x} P(x|u) \sum_{y} W^{\frac{1}{1+\rho}}(y|x)Q^{\frac{\rho}{1+\rho}}(y|u)  \Bigg]  \Bigg\},
\end{align}
where inequality (\ref{JENSEN1}) is due to Jensen's inequality. Next, we minimize the lower bound over $Q$. Differentiating the last expression w.r.t.\ $Q(y|u)$, we find that the minimizing distribution is given by
\begin{align}
\label{MINIQ1}
Q^{*}(y|u) = \frac{   \big[ \Phi(u,y,\rho) \big]^{1+\rho}  }{\sum_{y'} \big[ \Phi(u,y',\rho) \big]^{1+\rho} }. 
\end{align} 
Substituting (\ref{MINIQ1}) into (\ref{JENSEN1}), we get
\begin{align}
&-(1+\rho) \sum_{u} P(u)   
   \log \Bigg[ \sum_{x} P(x|u) \sum_{y} W^{\frac{1}{1+\rho}}(y|x)[Q^{*}(y|u)]^{\frac{\rho}{1+\rho}}  \Bigg] \nonumber \\
  &= -(1+\rho) \sum_{u} P(u)   
  \log  \left[ \sum_{x} P(x|u)  \sum_{y} W^{\frac{1}{1+\rho}}(y|x) \frac{   \big[ \Phi(u,y,\rho) \big]^{\rho}  }{\bigg\{  \sum_{y'} 
\big[ \Phi(u,y',\rho) \big]^{1+\rho} \bigg\}^{\frac{\rho}{1+\rho}}}  \right]   \\
  &= -(1+\rho) \sum_{u} P(u)   
   \log  \left[    \frac{ \sum_{y} \Big(  \Phi(u,y,\rho)  \cdot  \big[ \Phi(u,y,\rho) \big]^{\rho} \Big) }{\bigg\{  \sum_{y'} \big[ \Phi(u,y',\rho) \big]^{1+\rho} \bigg\}^{\frac{\rho}{1+\rho}}  }  \right] \\
  &= -(1+\rho) \sum_{u} P(u)   
   \log   \left[  \frac{ \sum_{y}  \big[ \Phi(u,y,\rho) \big]^{1+\rho}  }{\bigg\{  \sum_{y'} \big[ \Phi(u,y',\rho) \big]^{1+\rho} \bigg\}^{\frac{\rho}{1+\rho}}  }  \right]  \\
  &= -(1+\rho) \sum_{u} P(u)   
   \log  \left[  \bigg\{  \sum_{y} \big[ \Phi(u,y,\rho) \big]^{1+\rho} \bigg\}^{\frac{1}{1+\rho}} \right]   \\
 &= -\sum_{u} P(u)   
   \log  \left\{  \sum_{y} \big[ \Phi(u,y,\rho) \big]^{1+\rho}  \right\}  ,
\end{align}
which completes the proof of eq.\ (\ref{StrongTerm1}).  $\Box$

\subsection {Derivation of eq.\ (\ref{StrongTerm2})}

Similarly as in (\ref{ChangeOrder1})-(\ref{ChangeOrder2}),
\begin{align}
&\hat{E}_{y,2}( R_{y}, R_{z} )  \nonumber \\
&\overset{\bigtriangleup}{=} \min_{ V } \bigg\{  D( V \| W |  P   )  + \Big[ I(U;Y) + \Big[ I(X;Y|U)  -  R_{y} \Big]_{+} - R_{z} \Big]_{+}    \bigg\} \\
&=  \min_{ V } \Bigg\{  D( V \| W |  P   )  +  \max_{\mu \in [0,1] } \Bigg\{ \mu \cdot  \bigg[ I(U;Y) + \max_{\rho \in [0,1] } \Big\{ \rho \cdot  \big[ I(X;Y|U) - R_{y} \big] \Big\} - R_{z} \bigg] \Bigg\}    \Bigg\} \\
&=  \min_{ V } \Bigg\{  D( V \| W |  P   )  +  \max_{\mu \in [0,1] } \max_{\rho \in [0,1] }  \bigg\{ \mu \cdot  \Big[ I(U;Y) - R_{z} \Big]+   \mu \rho \cdot  \Big[ I(X;Y|U) - R_{y} \Big]   \bigg\}    \Bigg\} \\
&=  \min_{ V } \Bigg\{  D( V \| W |  P   )  +  \max_{\mu \in [0,1] } \max_{\lambda \in [0,\mu] }  \bigg\{ \mu \cdot  \Big[ I(U;Y) - R_{z} \Big]+   \lambda \cdot  \Big[ I(X;Y|U) - R_{y} \Big]   \bigg\}    \Bigg\} \\
&=  \max_{\mu \in [0,1] } \max_{\lambda \in [0,\mu] } \bigg\{ - \lambda R_{y} - \mu R_{z}  +  \min_{ V }  \Big\{  D( V \| W |  P   )  +  \mu \cdot  I(U;Y) + \lambda \cdot  I(X;Y|U)  \Big\}    \bigg\}.
\end{align}
Now, for the inner-most minimization,
\begin{align}
& \min_{ V }  \Big\{  D( V \| W |  P   )  +  \mu \cdot  I(U;Y) + \lambda \cdot  I(X;Y|U) \Big\}  \nonumber  \\
&= \min_{ V,Q,T }  \Bigg\{  \sum_{x,u}P(x,u)\sum_{y} V(y|x,u)\log \frac{V(y|x,u)}{W(y|x)}  \nonumber  \\  &\;\;\;\;\;\;\;\;\;\;\;\;\;  
+   \mu \sum_{x,u,y} P(x,u)V(y|x,u) \log \frac{Q(y|u)}{T(y)} + \lambda \sum_{x,u,y} P(x,u)V(y|x,u) \log \frac{V(y|x,u)}{Q(y|u)}  \Bigg\} \\
&= \min_{ V,Q,T }  \Bigg\{  \sum_{x,u,y} P(x,u)V(y|x,u) 
 \Bigg[  \log \frac{V(y|x,u)}{W(y|x)}+ \mu \log \frac{Q(y|u)}{T(y)} + \lambda \log \frac{V(y|x,u)}{Q(y|u)} \Bigg]  \Bigg\} \\
\label{BeforeMINIMUMV}
&= \min_{ V,Q,T }  \Bigg\{  \sum_{x,u,y} P(x,u)V(y|x,u) 
   \log \Bigg[ \frac{V^{1+\lambda}(y|x,u)}{W(y|x) T^{\mu}(y) Q^{\lambda - \mu}(y|u)}  \Bigg]  \Bigg\}.
\end{align}
First, we minimize over $V$. Holding $Q$ and $T$ fixed, and differentiating w.r.t.\ $V(y|x,u)$, we find that the minimizing $V$ is given by
\begin{align}
\label{MinimumV}
V^{*}(y|x,u) = \frac{ W^{\frac{1}{1+\lambda}}(y|x) T^{\frac{\mu}{1+\lambda}}(y) Q^{\frac{\lambda - \mu}{1+\lambda}}(y|u) }{\sum_{y'} W^{\frac{1}{1+\lambda}}(y'|x) T^{\frac{\mu}{1+\lambda}}(y') Q^{\frac{\lambda - \mu}{1+\lambda}}(y'|u) }. 
\end{align}
Substituting (\ref{MinimumV}) into (\ref{BeforeMINIMUMV}) and summing over $y$, we get
\begin{align}
& \min_{ Q,T }  \Bigg\{  \sum_{x,u,y} P(x,u)V^{*}(y|x,u) 
   \log \Bigg[ \frac{ [V^{*}(y|x,u)]^{1+\lambda} }{W(y|x) T^{\mu}(y) Q^{\lambda - \mu}(y|u)}  \Bigg]  \Bigg\} \nonumber  \\ 
&=  \min_{ Q,T }  \Bigg\{ -(1+ \lambda) \sum_{x,u} P(x,u) 
   \log \Bigg[ \sum_{y} W^{\frac{1}{1+\lambda}}(y|x) T^{\frac{\mu}{1+\lambda}}(y) Q^{\frac{\lambda - \mu}{1+\lambda}}(y|u)  \Bigg]  \Bigg\} \\
&=  \min_{ Q,T }  \Bigg\{ -(1+ \lambda) \sum_{u} P(u)  \sum_{x} P(x|u) 
   \log \Bigg[ \sum_{y} W^{\frac{1}{1+\lambda}}(y|x) T^{\frac{\mu}{1+\lambda}}(y) Q^{\frac{\lambda - \mu}{1+\lambda}}(y|u)  \Bigg]  \Bigg\} \\
\label{Jensen2}
&\geq  \min_{ Q,T }  \Bigg\{ -(1+ \lambda) \sum_{u} P(u)   
   \log \Bigg[ \sum_{x} P(x|u) \sum_{y} W^{\frac{1}{1+\lambda}}(y|x) T^{\frac{\mu}{1+\lambda}}(y) Q^{\frac{\lambda - \mu}{1+\lambda}}(y|u)  \Bigg]  \Bigg\},
\end{align}
where (\ref{Jensen2}) is due to Jensen's inequality. Next, we minimize the lower bound over $Q$, while holding $T$ fixed. Differentiating the last expression w.r.t.\ $Q(y|u)$, we find that the minimizing $Q$ is given by
\begin{align}
Q^{*}(y|u) = \frac{   \Big[ T^{\frac{\mu}{1+\lambda}}(y)  \sum_{x}  P(x|u)  W^{\frac{1}{1+\lambda}}(y|x) \Big]^{\frac{1+ \lambda}{1 + \mu}}  }{\sum_{y'}   \Big[ T^{\frac{\mu}{1+\lambda}}(y') \sum_{x'}  P(x'|u)  W^{\frac{1}{1+\lambda}}(y'|x') \Big]^{\frac{1+ \lambda}{1 + \mu}}  } . 
\end{align} 
Substituting into (\ref{Jensen2}), we have
\begin{align}
& \min_{ T }  \Bigg\{  -(1+ \lambda) \sum_{u} P(u)   
   \log \Bigg[ \sum_{x} P(x|u) \sum_{y} W^{\frac{1}{1+\lambda}}(y|x) T^{\frac{\mu}{1+\lambda}}(y) [Q^{*}(y|u)]^{\frac{\lambda - \mu}{1+\lambda}}  \Bigg]   \Bigg\}  \nonumber \\
&=  \min_{ T }   - (1+ \lambda) \sum_{u} P(u)   
   \log \left[ \sum_{x} P(x|u)  \right. \nonumber \\         &~~~~~~~~~~~~~~~~~~~~~~~ \left. \times  \sum_{y} W^{\frac{1}{1+\lambda}}(y|x) T^{\frac{\mu}{1+\lambda}}(y) \frac {   \Big[ T^{\frac{\mu}{1+\lambda}}(y)  \Phi(u,y,\lambda) \Big]^{\frac{1+ \lambda}{1 + \mu} \cdot {\frac{\lambda - \mu}{1+\lambda} } }  }{ \bigg\{ \sum_{y'}   \Big[ T^{\frac{\mu}{1+\lambda}}(y') \Phi(u,y',\lambda) \Big]^{\frac{1+ \lambda}{1 + \mu}} \bigg\} ^{\frac{\lambda - \mu}{1+\lambda}} }  \right] \\
  &= \min_{ T }  -(1+ \lambda) \sum_{u} P(u)    
   \log  \left[  \frac{ \sum_{y}  \Big[  T^{\frac{\mu}{1+\lambda}}(y)  \Phi(u,y,\lambda)  \Big] \cdot  \Big[  T^{\frac{\mu}{1+\lambda}}(y)  \Phi(u,y,\lambda)  \Big]^{ \frac{\lambda - \mu}{1+\mu} }  }{   \bigg\{ \sum_{y'}   \Big[ T^{\frac{\mu}{1+\lambda}}(y') \Phi(u,y',\lambda) \Big]^{\frac{1+ \lambda}{1 + \mu}} \bigg\} ^{\frac{\lambda - \mu}{1+\lambda}}  } \right]  \\
&= \min_{ T }  -(1+ \lambda) \sum_{u} P(u)   
   \log   \left[  \frac{ \sum_{y}    \Big[  T^{\frac{\mu}{1+\lambda}}(y)  \Phi(u,y,\lambda) \Big]^{ \frac{ 1 + \lambda }{1+\mu} }  }{   \bigg\{ \sum_{y'}   \Big[ T^{\frac{\mu}{1+\lambda}}(y') \Phi(u,y',\lambda) \Big]^{\frac{1+ \lambda}{1 + \mu}} \bigg\} ^{\frac{\lambda - \mu}{1+\lambda}}  } \right] \\
&= \min_{ T }  -(1+ \lambda) \sum_{u} P(u)   
   \log \left[    \Bigg\{ \sum_{y}   \Big[ T^{\frac{\mu}{1+\lambda}}(y) \Phi(u,y,\lambda) \Big]^{\frac{1+ \lambda}{1 + \mu}} \Bigg\} ^{\frac{ 1 + \mu }{1+\lambda}} \right]   \\
&= \min_{ T }  -(1+ \mu ) \sum_{u} P(u)   
   \log \left[  \sum_{y}   \Big[ T^{\frac{\mu}{1+\lambda}}(y) \Phi(u,y,\lambda) \Big]^{\frac{1+ \lambda}{1 + \mu}}  \right]    \\
&= \min_{ T }  -(1+ \mu ) \sum_{u} P(u)   
   \log \left[  \sum_{y}   T^{\frac{\mu}{1+\mu }}(y)  \big[ \Phi(u,y,\lambda) \big]^{\frac{1+ \lambda}{1 + \mu}}  \right]    \\
\label{Jensen3}
&\geq   \min_{ T }  -(1+ \mu )    
   \log \left[ \sum_{u} P(u)   \sum_{y}   T^{\frac{\mu}{1+\mu }}(y)  \big[  \Phi(u,y,\lambda) \big]^{\frac{1+ \lambda}{1 + \mu}} \right] .
\end{align}
Next, we minimize over $T$. Differentiating w.r.t.\ $T(y)$, we get
\begin{align}
T^{*}(y) = \frac{  \bigg\{  \sum_{u} P(u)  \big[   \Phi(u,y,\lambda) \big]^{\frac{1+ \lambda}{1 + \mu}} \bigg\}^{1 + \mu}  }{\sum_{y'}    \bigg\{  \sum_{u'} P(u')  \big[   \Phi(u',y',\lambda) \big]^{\frac{1+ \lambda}{1 + \mu}} \bigg\}^{1 + \mu}  } . 
\end{align} 
Substituting into (\ref{Jensen3}), we finally get
\begin{align}
& -(1+ \mu )    
   \log  \left[ \sum_{u} P(u)   \sum_{y}   [T^{*}(y)]^{\frac{\mu}{1+\mu }}  \big[  \Phi(u,y,\lambda) \big]^{\frac{1+ \lambda}{1 + \mu}} \right]  \nonumber \\
&=  -(1+ \mu )    
   \log  \left[ \sum_{u} P(u)   \sum_{y}   \big[  \Phi(u,y,\lambda) \big]^{\frac{1+ \lambda}{1 + \mu}}  
\right.       \nonumber  \\      & ~~~~~~~~~~~~~~~~~~~~~~~~~~~~~~~~~~~~~~~~~~~~   \left.  \times  
\frac{  \bigg\{  \sum_{\tilde{u}} P(\tilde{u})  \Big[   \Phi(\tilde{u},y,\lambda) \Big]^{\frac{1+ \lambda}{1 + \mu}} \bigg\}^{ \mu}  }{ \Bigg\{  \sum_{y'}    \bigg\{  \sum_{u'} P(u')  \Big[   \Phi(u',y',\lambda) \Big]^{\frac{1+ \lambda}{1 + \mu}} \bigg\}^{1 + \mu}   \Bigg\} ^{\frac{\mu}{1+\mu }} } \right] \\
&=  -(1+ \mu )    
   \log  \left[     \frac{ \sum_{y} \Bigg( \bigg\{  \sum_{u} P(u)  \Big[   \Phi(u,y,\lambda) \Big]^{\frac{1+ \lambda}{1 + \mu}} \bigg\}  \cdot  \bigg\{  \sum_{\tilde{u}} P(\tilde{u})  \Big[  \Phi(\tilde{u},y,\lambda) \Big]^{\frac{1+ \lambda}{1 + \mu}} \bigg\}^{ \mu} \Bigg) }{ \Bigg\{  \sum_{y'}    \bigg\{  \sum_{u'} P(u')  \Big[ \Phi(u',y',\lambda) \Big]^{\frac{1+ \lambda}{1 + \mu}} \bigg\}^{1 + \mu}   \Bigg\} ^{\frac{\mu}{1+\mu }} } \right] \\
&=  -(1+ \mu )    
   \log    \left[   \frac{ \sum_{y}   \bigg\{  \sum_{u} P(u)  \Big[   \Phi(u,y,\lambda) \Big]^{\frac{1+ \lambda}{1 + \mu}} \bigg\}^{1 +  \mu}  }{ \Bigg\{  \sum_{y'}    \bigg\{  \sum_{u'} P(u')  \Big[  \Phi(u',y',\lambda) \Big]^{\frac{1+ \lambda}{1 + \mu}} \bigg\}^{1 + \mu}   \Bigg\} ^{\frac{\mu}{1+\mu }} } \right] \\
&=  -(1+ \mu )    
   \log      \Bigg\{  \sum_{y}    \Bigg\{  \sum_{u} P(u)  \big[  \Phi(u,y,\lambda) \big]^{\frac{1+ \lambda}{1 + \mu}} \Bigg\}^{1 + \mu}   \Bigg\} ^{\frac{1}{1+\mu }}  \\
&=     
 -  \log      \sum_{y}    \Bigg\{  \sum_{u} P(u)  \big[  \Phi(u,y,\lambda) \big]^{\frac{1+ \lambda}{1 + \mu}} \Bigg\}^{1 + \mu}.
\end{align} 
Hence, (\ref{StrongTerm2}) is now proved, as well as the lower bound given in (\ref{LowerBoundWeak}). $\Box$

\section{Analyzing the Gallager-Style Lower Bounds}

\subsection{A Study for $E_{y,1}(R_{y})$}
As in the single user case, we expect to find a critical rate and a maximal rate. By maximal rate, that will be denoted by $R_{\max}$, 
we mean $\sup \{R_{y}:~  E_{y,1}(R_{y})>0 \}$. 
By critical rate, to be denoted by $R_{\mbox{\tiny crit}}$, we mean the boundary between the range where $E_{y,1}(R_{y})$ is affine and the range where it is curvy.
\\
Let 
\begin{align}
\label{DM0}
E_{y,1}(R_{y}) = \max_{\rho \in [0,1] } \Big\{ E_{0}(\rho)  - \rho R_{y}   \Big\} ,
\end{align}
where we have defined
\begin{align}
E_{0}(\rho) =  - \sum_{u} P(u) \log \sum_{y} \bigg[ \sum_{x} P(x|u)W^{\frac{1}{1+\rho}}(y|x)     \bigg]^{1+\rho}.
\end{align}
Setting the partial derivative of the bracketed part of (\ref{DM0}) equal to 0, we get
\begin{align}
\label{SolutionforRy}
R_{y} =  \frac{\partial E_{0}(\rho)}{\partial \rho} .        
\end{align}
Following the same considerations as in \cite[Section 5.6]{GAL68}, if some $\rho \in [0,1]$ satisfies (\ref{SolutionforRy}), then it must maximize (\ref{DM0}). It turns out that a solution to (\ref{SolutionforRy}) exists if
\begin{align}
\label{RyRange}
\left.\frac{\partial E_{0}(\rho)}{\partial \rho} \right |_{\rho=1} \leq  R_{y} \leq  \left.\frac{\partial E_{0}(\rho)}{\partial \rho} \right |_{\rho=0} .        
\end{align}
In this range, it is convenient to use (\ref{SolutionforRy}) to relate $E_{y,1}(R_{y})$ and $R_{y}$ parametrically as functions of $\rho$. For the interval $0 \leq \rho \leq 1$, this gives
\begin{align}
E_{y,1}(R_{y}) &=   E_{0}(\rho)  - \rho \cdot \frac{\partial E_{0}(\rho)}{\partial \rho}    ,  \\
R_{y} &=  \frac{\partial E_{0}(\rho)}{\partial \rho} . 
\end{align}
For $R_{y} < \partial E_{0}(\rho) / \partial \rho |_{\rho = 1}  $, the parametric equations are not valid. In this case, the maximum occurs at $\rho = 1$. Thus, $E_{y,1}(R_{y})$ is affine with slope $-1$:
\begin{align}
E_{y,1}(R_{y}) &=   E_{0}(1)  -  R_{y},
\end{align}
where 
\begin{align}
E_{0}(1) =  - \sum_{u} P(u) \log \sum_{y} \bigg[ \sum_{x} P(x|u) \sqrt { W(y|x) }    \bigg]^{2}.
\end{align}
Now, we can find $R_{\max}$ and $R_{\mbox{\tiny crit}}$, which are given by the right-most side and the left-most side of (\ref{RyRange}), respectively. Differentiating $E_{0}(\rho)$ w.r.t.\ $\rho$ and substituting $\rho = 0$ gives 
\begin{align}
R_{\max} =  \left.\frac{\partial E_{0}(\rho)}{\partial \rho} \right |_{\rho=0}  =  I_{P,W}(X;Y|U),       
\end{align}
where $I_{P,W}(X;Y|U)$ is the conditional mutual information induced by the channel $W(y|x)$ and the code distribution $P(u,x)$. Next, define 
\begin{align}
F(u,y) =   \sum_{x} P(x|u) \sqrt { W(y|x) }    ,
\end{align}
and 
\begin{align}
G(u,y) =   \sum_{x} P(x|u) \sqrt { W(y|x) } \log W(y|x)   .
\end{align}
After some algebra, we find that
\begin{align}
R_{\mbox{\tiny crit}} =  \left.\frac{\partial E_{0}(\rho)}{\partial \rho} \right |_{\rho=1}  = - \sum_{u} P(u) \frac{\sum_{y} \Big[ F^{2}(u,y) \log F(u,y) - \frac{1}{2} F(u,y) G(u,y)      \Big]}{ \sum_{y'}  F^{2}(u,y') } .     
\end{align}

\subsection{A Study for $E_{y,2}(R_{y}, R_{z})$}
Let 
\begin{align}
E_{y,2}(R_{y}, R_{z})  
&= \max_{\mu \in [0,1] } \max_{\lambda \in [0,\mu] } \Bigg\{   -  \log \sum_{y}  \Bigg(  \sum_{u} P(u)  \bigg[ \sum_{x} P(x|u)W^{\frac{1}{1+\lambda}}(y|x)   \bigg]^{\frac{1+\lambda}{1+\mu}}  \Bigg)^{1+\mu}  \nonumber \\
&~~~~~~~~~~~~~~~~~~~~~~~~~~~~~~~~~~~~~~~~~~~
~~~~~~~~~~~~~~~~~~~  - \lambda R_{y}  -\mu R_{z}   \Bigg\} \\
&= \max_{\mu \in [0,1] } \max_{s \in [0,1] } \Bigg\{  -  \log \sum_{y}  \Bigg(  \sum_{u} P(u)  \bigg[ \sum_{x} P(x|u)W^{\frac{1}{1+ s \mu}}(y|x)   \bigg]^{\frac{1+ s \mu}{1+\mu}}  \Bigg)^{1+\mu}  \nonumber \\
&~~~~~~~~~~~~~~~~~~~~~~~~~~~~~~~~~~~~~~~~~~~
~~~~~~~~~~~~~~~~~~~ - s \mu R_{y}  -\mu R_{z}   \Bigg\} ,
\end{align}
and define
\begin{align}
E_{1}(s, \mu) &=  -  \log \sum_{y}  \Bigg\{  \sum_{u} P(u)  \bigg[ \sum_{x} P(x|u)W^{\frac{1}{1+s \mu}}(y|x)   \bigg]^{\frac{1+s \mu}{1+\mu}}  \Bigg\}^{1+\mu},
\end{align}
such that
\begin{align}
\label{DeriveMe}
E_{y,2}(R_{y}, R_{z})  
= \max_{\mu \in [0,1] } \max_{s \in [0,1] } \Big\{ E_{1}(s, \mu)  - s \mu R_{y}  -\mu R_{z}   \Big\} .
\end{align}
Setting the partial derivatives of the bracketed part of (\ref{DeriveMe}) to zero, we get
\begin{align}
\label{DM1}
   \frac{\partial }{\partial s}  E_{1}(s, \mu )  &=   \mu R_{y}     \\ 
\label{DM2}
   \frac{\partial }{\partial \mu}  E_{1}(s, \mu )  &= s R_{y} + R_{z}  ,
\end{align}
or, equivalently,
\begin{align}
R_{y} &= \frac{1}{\mu} \cdot  \frac{\partial }{\partial s}  E_{1}(s, \mu)     \\ 
R_{z} &=  \frac{\partial }{\partial \mu}  E_{1}(s, \mu)  -  \frac{s}{\mu} \cdot  \frac{\partial }{\partial s}  E_{1}(s, \mu)  .
\end{align}
Now, if some $(\mu, s) \in [0,1]^{2}$ satisfies (\ref{DM1}) and (\ref{DM2}), we may relate $E_{y,2}(R_{y}, R_{z})$, $R_{y}$ and $R_{z}$ parametrically as functions of $s$ and $\mu$. This gives
\begin{align}
\label{GeneralRep1}
E_{y,2}(R_{y}, R_{z}) &=  E_{1}(s, \mu)  -  \mu  \cdot  \frac{\partial }{\partial \mu}   E_{1}(s, \mu)    \\
R_{y} &=  \frac{1}{\mu}  \cdot  \frac{\partial   }{\partial s}  E_{1}(s, \mu)        \\ 
\label{GeneralRep2}
R_{z} &=  \frac{\partial   }{\partial \mu}   E_{1}(s, \mu)  -   \frac{s}{\mu}  \cdot  \frac{\partial   }{\partial s}  E_{1}(s, \mu) .
\end{align}
For explicit expressions for the partial derivatives of $E_{1}(s, \mu)$ w.r.t.\ $s$ and $\mu$, we first define 
\begin{align}
A(y,s,\mu) &=   \sum_{u} P(u)  \bigg[ \sum_{x} P(x|u)W^{\frac{1}{1+s \mu}}(y|x)   \bigg]^{\frac{1+s \mu}{1+\mu}}    \\ 
B(u,y,s,\mu) &=   \sum_{x} P(x|u)W^{\frac{1}{1+s \mu}}(y|x)       \\
E(u,y,s,\mu) &=   \sum_{x} P(x|u)W^{\frac{1}{1+s \mu}}(y|x)  \log W(y|x)       \\
C(y,s,\mu) &=   \sum_{u} P(u)  \big[  B(u,y,s,\mu)  \big]^{\frac{1+s \mu}{1+\mu}} \Bigg( \frac{s-1}{1+\mu} \cdot  \log B(u,y,s,\mu) - \frac{s}{1+s \mu} \cdot \frac{E(u,y,s,\mu)}{B(u,y,s,\mu)}  \Bigg)  \\
D(y,s,\mu) &=   \sum_{u} P(u)  \big[  B(u,y,s,\mu)  \big]^{\frac{1+s \mu}{1+\mu}} \Bigg(\log B(u,y,s,\mu) - \frac{1}{1+s \mu} \cdot \frac{E(u,y,s,\mu)}{B(u,y,s,\mu)}  \Bigg),
\end{align}
and get that the partial derivative w.r.t.\ $s$ is given by
\begin{align}
 \frac{\partial }{\partial s}  E_{1}(s, \mu) =  - \mu \cdot \frac{\sum_{y} A^{\mu}(y,s,\mu)D(y,s,\mu) }{\sum_{y'} A^{1+\mu}(y',s,\mu)} ,
\end{align}
and the partial derivatives w.r.t.\ $\mu$ is given by
\begin{align}
 \frac{\partial }{\partial \mu}  E_{1}(s, \mu) =  -  \frac{\sum_{y} A^{1+\mu}(y,s,\mu) \cdot \Big[ \log A(y,s,\mu) + \frac{C(y,s,\mu)}{A(y,s,\mu)}    \Big] }{\sum_{y'} A^{1+\mu}(y',s,\mu)} .
\end{align}
Consider the rate pair $(R_{y},R_{z})$ for which both (\ref{DM1}) and (\ref{DM2}) hold with $s=\mu=1$:
\begin{align}
R_{y} &=  \left. \frac{\partial }{\partial s}  E_{1}(s, 1) \right |_{s=1}        \\ 
R_{y} + R_{z} &=  \left.  \frac{\partial }{\partial \mu}  E_{1}(1, \mu)  \right |_{\mu=1},
\end{align}
which is the corner point of the affine rate region. As an immediate consequence, due to the monotonicity of $E_{1}$, we get that for low rates, i.e., for rate pairs $(R_{y},R_{z})$ that satisfy 
\begin{align}
R_{y}  \leq    \left. \frac{\partial }{\partial s}  E_{1}(s, 1) \right |_{s=1}   ,   
\end{align}
and
\begin{align}
\label{SumRate}
 R_{y} + R_{z} \leq  \left. \frac{\partial }{\partial \mu}  E_{1}(1, \mu) \right |_{\mu=1},  
\end{align} 
the maximizers are $s^{*}=\mu^{*}=1$, and $E_{y,2}(R_{y}, R_{z})$ is given by
\begin{align}
E_{y,2}(R_{y}, R_{z}) =   E_{1}(1, 1)  - ( R_{y} +  R_{z} )  ,
\end{align}
where
\begin{align}
E_{1}(1, 1)  =  -  \log \sum_{y}  \Bigg\{  \sum_{x} P(x) \sqrt{W(y|x)}  \Bigg\}^{2}.
\end{align}
According to (\ref{SumRate}), we can find the maximal sum-rate in the affine region. Let 
\begin{align}
F(y) =   \sum_{x} P(x) \sqrt { W(y|x) }    ,
\end{align}
and 
\begin{align}
G(y) =   \sum_{x} P(x) \sqrt { W(y|x) } \log W(y|x)   .
\end{align}
After some algebra, we find that the maximal sum-rate is given by
\begin{align}
R_{y} + R_{z} \leq  \left. \frac{\partial }{\partial \mu}  E_{1}(1, \mu) \right |_{\mu=1}   = -  \frac{\sum_{y} \Big[ F^{2}(y) \log F(y) - \frac{1}{2} F(y) G(y)      \Big]}{ \sum_{y'}  F^{2}(y') } .
\end{align}
\\
The error exponent $E_{y,2}(R_{y}, R_{z})$ depends solely on $R_{z}+R_{y}$ if and only if the maximizing $s \in [0,1]$ is $s^{*}=1$. In this case,
\begin{align}
&E_{y,2}(R_{y}, R_{z}) \nonumber  \\
&= \max_{\mu \in [0,1] }  \Bigg\{  -  \log \sum_{y}  \Bigg\{  \sum_{u} P(u)  \bigg[ \sum_{x} P(x|u)W^{\frac{1}{1+ \mu}}(y|x)   \bigg]^{\frac{1+ \mu}{1+\mu}}  \Bigg\}^{1+\mu}    -  \mu R_{y}  -\mu R_{z}    \Bigg\} \\
&= \max_{\mu \in [0,1] }  \Bigg\{  -  \log \sum_{y}  \Bigg\{  \sum_{u} P(u)  \sum_{x} P(x|u)W^{\frac{1}{1+ \mu}}(y|x)  \Bigg\}^{1+\mu}    -  \mu ( R_{y}  +  R_{z} )   \Bigg\} \\
\label{SumRateResult}
&= \max_{\mu \in [0,1] }  \Bigg\{  -  \log \sum_{y}  \Bigg\{  \sum_{x} P(x) W^{\frac{1}{1+ \mu}}(y|x)  \Bigg\}^{1+\mu}    -  \mu ( R_{y}  +  R_{z} )   \Bigg\},
\end{align}
which means that $E_{y,2}(R_{y}, R_{z}) = E_{\mbox{\tiny r}}(R_{y}+ R_{z}, P_{X})$, i.e., the ordinary random coding error exponent at rate $R_{y}  +  R_{z}$ for an i.i.d.\ drawn code with distribution $P_{X}$. Next, we find the rate region for which $E_{y,2}(R_{y}, R_{z})$ depends solely on $R_{z}+R_{y}$, but it is not affine in the sum-rate. 
Consider the rate pairs $(R_{y},R_{z})$ for which both (\ref{DM1}) and (\ref{DM2}) hold with $s=1$:
\begin{align}
\label{S1-eq1}
 \mu R_{y} &=  \left. \frac{\partial }{\partial s}  E_{1}(s, \mu) \right |_{s=1}        \\ 
\label{S1-eq2}
  R_{y} + R_{z} &=   \frac{\partial }{\partial \mu}  E_{1}(1, \mu)    .
\end{align}
Let $\Gamma_{S1}$ denote the curve given by eqs.\ (\ref{S1-eq1})-(\ref{S1-eq2}):
\begin{align}
\Gamma_{S1} = \Bigg\{  (R_{y},R_{z}) \Bigg|   R_{y} =  \frac{1}{\mu} \cdot  \left. \frac{\partial }{\partial s}  E_{1}(s, \mu) \right |_{s=1}, \;\;\;  R_{y} + R_{z} =   \frac{\partial }{\partial \mu}  E_{1}(1, \mu), \;\;\;  0< \mu \leq 1      \Bigg\}.
\end{align}
Now, in the set of $(R_{y},R_{z})$, with 
\begin{align}
\label{SumRateInequality}
\left. \frac{\partial }{\partial \mu}  E_{1}(1, \mu) \right |_{\mu=1} \leq  R_{y} + R_{z} \leq  \left. \frac{\partial }{\partial \mu}  E_{1}(1, \mu) \right |_{\mu=0}    ,
\end{align}
and being underneath the curve $\Gamma_{S1}$, the maximizer is $s^{*}=1$, and $E_{y,2}(R_{y}, R_{z})$ is given by (\ref{SumRateResult}). Notice that the left-hand-side and the right-hand-side of (\ref{SumRateInequality}) are expressions for the critical-rate and the maximal-rate for the channel $W_{1}(y|x)$, respectively, and hence, the latter cannot be smaller than the former.   
Before moving forward, let us obtain a simple information-theoretic expression for the maximal sum-rate. According to the right-hand side of (\ref{SumRateInequality}), we only have to differentiate $E_{1}(1, \mu)$ w.r.t.\ $\mu$ and then substitute $\mu=0$. We get
\begin{align}
R_{y} + R_{z} \leq  \left. \frac{\partial }{\partial \mu}  E_{1}(1, \mu) \right |_{\mu=0} =  I_{P_{X},W}(X;Y)   ,
\end{align}
where $I_{P_{X},W}(X;Y)$ is the mutual information induced by the channel $W_{1}(y|x)$ and the code distribution $P(x)$. 

Let us now turn to the other extreme, where $E_{y,2}(R_{y}, R_{z})$ depends solely on $R_{z}$. This happens if and only if the maximizing $s \in [0,1]$ is $s^{*}=0$. In this case,
\begin{align}
E_{y,2}(R_{y}, R_{z}) 
&= \max_{\mu \in [0,1] } \Bigg\{  -  \log \sum_{y}  \Bigg\{  \sum_{u} P(u)  \bigg[ \sum_{x} P(x|u)W(y|x)   \bigg]^{\frac{1}{1+\mu}}  \Bigg\}^{1+\mu}     -\mu R_{z}    \Bigg\} \\
&= \max_{\mu \in [0,1] } \Bigg\{  -  \log \sum_{y}  \Bigg\{  \sum_{u} P(u)  V^{\frac{1}{1+\mu}}(y|u)    \Bigg\}^{1+\mu}     -\mu R_{z}    \Bigg\},
\end{align}
which means that $E_{y,2}(R_{y}, R_{z}) = E_{\mbox{\tiny r}}(R_{z}, P_{U})$, i.e., the ordinary random coding error exponent at rate $R_{z}$ for an i.i.d.\ code with distribution $P_{U}$, where $V$ is defined to be the equivalent channel from $U$ to $Y$. The simple explanation for the fact that $E_{y,2}(R_{y}, R_{z})$ becomes independent of $R_{y}$, for high $R_{y}$, is that the satellite codewords behave like pure noise. 

Next, we find the region where $E_{y,2}(R_{y}, R_{z})$ depends solely on $R_{z}$. 
Consider the rate pairs $(R_{y},R_{z})$ for which both (\ref{DM1}) and (\ref{DM2}) hold with $s=0$:
\begin{align}
\label{S0-eq1}
 \mu R_{y} &=  \left. \frac{\partial }{\partial s}  E_{1}(s, \mu) \right |_{s=0}        \\ 
\label{S0-eq2}
   R_{z} &=   \frac{\partial }{\partial \mu}  E_{1}(0, \mu)    ,
\end{align}
Let $\Gamma_{S0}$ denote the curve given by eqs.\ (\ref{S0-eq1})-(\ref{S0-eq2}):
\begin{align}
\label{S0_Curve}
\Gamma_{S0} = \Bigg\{  (R_{y},R_{z}) \Bigg|   R_{y} =  \frac{1}{\mu} \cdot  \left. \frac{\partial }{\partial s}  E_{1}(s, \mu) \right |_{s=0}, \;\;\;  R_{z} =   \frac{\partial }{\partial \mu}  E_{1}(0, \mu), \;\;\;   0< \mu \leq 1      \Bigg\}.
\end{align}
In addition, we have the following corner point for $\mu = 1$: 
\begin{align}
(\tilde{R}_{y},\tilde{R}_{z}) = \Bigg(     \left. \frac{\partial }{\partial s}  E_{1}(s, 1) \right |_{s=0},   
       \left.  \frac{\partial }{\partial \mu}  E_{1}(0, \mu) \right |_{\mu = 1}   \Bigg),
\end{align}
and we use it to define the straight line connecting that corner point to the $R_{y}$-axis: 
\begin{align}
\tilde{\Gamma}_{S0} = \Bigg\{  (R_{y},R_{z}) \Bigg|   R_{y} =  \tilde{R}_{y},   0 \leq  R_{z}  \leq  \tilde{R}_{z}     \Bigg\},
\end{align}
which is the set of all $(R_{y},R_{z})$ for which the maximizers are $s^{*}=0$ and $\mu^{*}=1$. \\
\\
Let $\hat{\Gamma}_{S0}$ be defined by $\Gamma_{S0} \cup \tilde{\Gamma}_{S0}$. In fact, the curve $\hat{\Gamma}_{S0}$ is the borderline between the region where $E_{y,2}(R_{y}, R_{z})$ depends on $R_{y}$ (affine or curvy) and the region where $E_{y,2}(R_{y}, R_{z})$ is independent of $R_{y}$. The set of all $(R_{y},R_{z})$ that are above the curve $\hat{\Gamma}_{S0}$ defines the region where $E_{y,2}(R_{y}, R_{z})$ is independent of $R_{y}$. In addition, let us obtain a simple informational expression for the maximum of $R_{z}$. According to (\ref{S0_Curve}), we only have to differentiate $E_{1}(0, \mu)$ w.r.t.\ $\mu$ and then substitute $\mu=0$. We get
\begin{align}
 R_{z} \leq  \left. \frac{\partial }{\partial \mu}  E_{1}(0, \mu) \right |_{\mu=0} =  I_{P_{U},V}(U;Y)   ,
\end{align}
where $I_{P_{U},V}(U;Y)$ is the mutual information induced by the channel $\{V(y|u)\}$ and $\{P(u)\}$. In the region 
\begin{align}
R_{z} \leq \left. \frac{\partial }{\partial \mu}  E_{1}(0, \mu) \right |_{\mu=1} ,
\end{align}
the maximizer is $\mu^{*} = 1$, and $E_{y,2}(R_{y}, R_{z})$ is affine in $R_{z}$ and is given by
\begin{align}
E_{y,2}(R_{y}, R_{z}) =   E_{1}(0, 1)  -  R_{z}   ,
\end{align}
where
\begin{align}
E_{1}(0, 1)  =  -  \log \sum_{y}  \Bigg\{  \sum_{u} P(u) \sqrt{V(y|u)}  \Bigg\}^{2}.
\end{align}
\\
The third region is the set of all $(R_{y},R_{z})$ for which the maximizing $s$ is in $(0,1)$. In this case, we use (\ref{GeneralRep1})-(\ref{GeneralRep2}), which hold for every $s \in (0,1)$ and $\mu \in [0,1]$ such that both (\ref{DM1}) and (\ref{DM2}) are satisfied.
This region can be devided into two complementary regions. In the first one, the maximizer is $\mu^{*} = 1$, and $E_{y,2}(R_{y}, R_{z})$ is affine in $R_{z}$ and curvy in $R_{y}$, while in the second one, the maximizer $\mu^{*}$ is in $(0,1)$, and $E_{y,2}(R_{y}, R_{z})$ is curvy in both $R_{z}$ and $R_{y}$. The borderline between those two regions is given by the curve
\begin{align}
\Gamma_{\mu 1} = \Bigg\{  (R_{y},R_{z}) \Bigg|   R_{y} =  \frac{\partial }{\partial s}  E_{1}(s, 1) , \;\;\;    sR_{y} + R_{z} =  \left. \frac{\partial }{\partial \mu}  E_{1}(s, \mu) \right |_{\mu=1}  , \;\;\;  0 \leq s \leq 1      \Bigg\}.
\end{align}
For 
\begin{align}
 R_{z} &\geq  I_{P_{U},V}(U;Y)   \nonumber  \\
 R_{y} + R_{z} &\geq  I_{P_{X},W}(X;Y)   ,
\end{align}
the maximizers are $s^{*}=\mu^{*}=0$, and then $E_{y,2}(R_{y}, R_{z}) = 0$.

\section*{Appendix}
\renewcommand{\theequation}{A.\arabic{equation}}
    \setcounter{equation}{0}
\subsection*{Proof of Theorem 3}

Regarding decoder (\ref{GeneralDecoder})-(\ref{GeneralDecoder2}), let us select
\begin{align}
\label{AgoodFunction2}
  f(Q_{UXZ}) =   I_{Q}(U;Z)  +  I_{Q}(X;Z|U) . 
\end{align}
We show that (\ref{AgoodFunction2}) achieves the maximum of $ E_{2}(s_{1}(Q_{U_{0}X_{00}Z}), Q_{Z})$, given by Lemma 3, and therefore, the error exponent of this decoder is as large as that of the optimal decoder. First, the threshold $s_{0}(Q_{U_{0}Z}) $ can be easily simplified as
\begin{align}
 s_{0}(Q_{U_{0}Z})   &=  R_{y} + \max_{\big\{ Q_{X|U_{0}Z} \in \mathcal{S}(Q_{U_{0}Z}): ~ I_{Q}(X;Z|U_{0}) \leq R_{y} \big\} } \Big[ f(Q_{U_{0}XZ}) - I_{Q}(X;Z|U_{0}) \Big] \\
&=  R_{y} + \max_{\big\{ Q_{X|U_{0}Z} \in \mathcal{S}(Q_{U_{0}Z}): ~ I_{Q}(X;Z|U_{0})  \leq R_{y} \big\} } \Big[ I_{Q}(U_{0};Z)  +  I_{Q}(X;Z|U_{0})   - I_{Q}(X;Z|U_{0}) \Big]  \\
&=  R_{y} +  I_{Q}(U_{0};Z) ,
\end{align}  
such that 
\begin{align}
s_{1}(Q_{U_{0}X_{00}Z}) &= 
 \max \Big\{ R_{y} +  I_{Q}(U_{0};Z),  I_{Q}(U_{0};Z)  +   I_{Q}(X_{00};Z|U_{0})  \Big\} \\
&=  I_{Q}(U_{0};Z)  +   \max \Big\{ R_{y} ,  I_{Q}(X_{00};Z|U_{0})  \Big\}.
\end{align}
In general, the constraint of the inner minimization problem defining $E_{2}(s_{1}(Q_{U_{0}X_{00}Z}), Q_{Z})$ is given by
\begin{align}
f(Q_{UXZ}) +\big[ R_{y} - I_{Q}(X;Z|U) \big]_{+} \geq s_{1}(Q_{U_{0}X_{00}Z}),
\end{align}
which can now be written as
\begin{align}
&I_{Q}(U;Z)  +   I_{Q}(X;Z|U) +\big[ R_{y}- I_{Q}(X;Z|U)  \big]_{+} \nonumber 
\\ &~~~~~~~~~~~~~~~~~~~~~~~~~~~~~~\geq I_{Q}(U_{0};Z)  +   \max \Big\{ R_{y} ,  I_{Q}(X_{00};Z|U_{0})  \Big\}.
\end{align}
Substracting $R_{y}$ from both sides gives
\begin{align}
&I_{Q}(U;Z)  +   I_{Q}(X;Z|U)  -  R_{y}  +\big[ R_{y}- I_{Q}(X;Z|U)  \big]_{+} \nonumber 
\\ &~~~~~~~~~~~~~~~~~~~~~~~~~~~~~~\geq I_{Q}(U_{0};Z)  +   \max \Big\{ 0 ,  I_{Q}(X_{00};Z|U_{0})  -  R_{y} \Big\},  
\end{align}
or, 
\begin{align}
I_{Q}(U;Z) + \big[ I_{Q}(X;Z|U) - R_{y} \big]_{+}  \geq  I_{Q}(U_{0};Z)  +  \big[  I_{Q}(X_{00};Z|U_{0})  -  R_{y} \big]_{+}.  
\end{align}
Defining $D(Q_{UXZ}) \overset{\bigtriangleup}{=} I_{Q}(U;Z) + \big[ I_{Q}(X;Z|U) - R_{y} \big]_{+}$, we have
\begin{align}
  E_{2}(s_{1}(Q_{U_{0}X_{00}Z}), Q_{Z}) 
&=   \min_{ Q_{UX|Z} \in \mathcal{S}(Q_{Z}): ~ D(Q_{UXZ})  \geq D(Q_{U_{0}X_{00}Z}) }    D(Q_{UXZ})    \\
&= I_{Q}(U_{0};Z)  +  \big[  I_{Q}(X_{00};Z|U_{0})  -  R_{y} \big]_{+} ,
\end{align}
which is the same as on the right hand side of (\ref{Lemma2eq}). $\Box$

\end{document}